\documentclass[a4paper,11pt]{article}
\pdfoutput=1 

\usepackage{jcappub} 

\usepackage[T1]{fontenc} 
\usepackage{xcolor}
\usepackage{graphicx}
\usepackage{mwe}
\usepackage{physics}
\usepackage{caption}
\usepackage{subcaption}
\usepackage{titlesec}

\title{Deconstructing the neutrino mass constraint from galaxy redshift surveys}

\author[a,1]{Aoife Boyle\note{Corresponding author.}}
\author[a,b]{and Eiichiro Komatsu}

\affiliation[a]{Max-Planck-Institut f\"{u}r Astrophysik, Karl-Schwarzschild-Str. 1, 85748 Garching bei M\"{u}nchen, Germany}
\affiliation[b]{Kavli Institute for the Physics and Mathematics of the Universe, Todai Institutes for Advanced Study, the University of Tokyo, Kashiwa, Japan 277-8583 (Kavli IPMU, WPI}

\emailAdd{aoife@mpa-garching.mpg.de, komatsu@mpa-garching.mpg.de}

\abstract{The total mass of neutrinos can be constrained in a number of ways using galaxy redshift surveys. Massive neutrinos modify the expansion rate of the Universe, which can be measured using baryon acoustic oscillations (BAOs) or the Alcock-Paczynski (AP) test. Massive neutrinos also change the structure growth rate and the amplitude of the matter power spectrum, which can be measured using redshift-space distortions (RSD). We use the Fisher matrix formalism to disentangle these information sources, to provide projected neutrino mass constraints from each of these probes alone and to determine how sensitive each is to the assumed cosmological model. We isolate the distinctive effect of neutrino free-streaming on the matter power spectrum and structure growth rate as a signal unique to massive neutrinos that can provide the most robust constraints, which are relatively insensitive to extensions to the cosmological model beyond $\Lambda$CDM. We also provide forecasted constraints using all of the information contained in the observed galaxy power spectrum combined, and show that these maximally optimistic constraints are primarily limited by the accuracy to which the optical depth of the cosmic microwave background, $\tau$, is known.}

\begin{document}
\maketitle
\flushbottom

\section{Introduction}
Neutrinos have at least three individual mass states, but direct experimental measurements of their absolute mass values have so far proved impossible because of their incredibly small values. A range of neutrino oscillation experiments have allowed measurements of the mass differences between the three species and the corresponding minimum total mass \cite{capozzi_neutrino_2016}. While particle physics experiments have provided the lower bound on the total neutrino mass, $M_{\nu}$, the current tentative upper limit comes from cosmological studies. As massive neutrinos constitute a small fraction of the energy density of the universe, a range of cosmological probes can provide indirect evidence of their mass properties. 

The Planck survey set a 95\% upper limit of $\sim$230 meV on the total neutrino mass using data from the cosmic microwave background (CMB) for a flat $\Lambda$CDM cosmological model extended to include massive neutrinos, using a dataset combining CMB temperature anisotropies and lensing, plus external large-scale structure probes (a dataset referred to as \texttt{TT+lensing+ext} in the article) \cite{planck_collaboration_planck_2016}. It is expected that next-generation CMB surveys \cite{abazajian_cmb-s4_2016} will enhance this constraint further. Meanwhile, galaxy redshift surveys provide a strong complement to CMB information, in the form of three-dimensional information on galaxy clustering patterns that allow us to trace the evolution of structure in the universe over time \cite{hu_weighing_1998, takada_cosmology_2006}. It has been predicted that the combination of CMB and large-scale structure surveys will allow for a constraint on the neutrino mass at the tens-of-meV level in the coming decade \cite{wu_guide_2014, abazajian_neutrino_2015, allison_towards_2015, pan_constraints_2015, archidiacono_physical_2017}.

The galaxy power spectrum measured by galaxy redshift surveys provides multiple probes that can be used to constrain cosmological parameters. Massive neutrinos alter cosmological distances through their modification of the expansion rate, and cosmological distance scales can be constrained using standard rulers such as the baryon acoustic oscillation (BAO) scale \cite{eisenstein_detection_2005, cole_2df_2005}, and through the Alcock-Paczynski (AP) test \cite{alcock_evolution_1979}, which requires that isotropy is conserved when models are converted from redshift space. BAO information, in particular, is a popular probe of cosmological parameters because it is easily understood using linear theory and is easily measured, and does not depend heavily on an understanding of galaxy bias. For this reason, many previous studies have focused on extracting neutrino mass constraints from this source alone \cite{pan_constraints_2015, allison_towards_2015}. Redshift-space distortions (RSD) \cite{kaiser_clustering_1987} are used to constrain the growth of structure and are also affected by the additional matter provided by non-relativistic neutrinos. 

However, there is nothing unique about the qualitative effect of massive neutrinos on the expansion rate of the universe or the large-scale structure growth rate. Either of these effects could be mimicked by the addition of other kinds of matter, or by changes in the nature of dark energy. The aim of this paper is to deconstruct the information used to constrain the neutrino mass from galaxy redshift surveys, and to isolate the information available from those signals that are uniquely identifiable as the effects of massive neutrinos.

It is well known that neutrino free-streaming suppresses the growth of structure on small scales relative to that on large scales to an extent that is proportional to their mass \cite[for a thorough review, see][]{lesgourgues_massive_2006}. This results in small but distinctive signatures in the matter power spectrum $P_{\textrm{m}}(k,z)$ and in the structure growth rate $f(k, z)$. The possibility of constraining the neutrino mass through a scale-dependent measurement of $f(k,z)$ from RSD was recently explored by \cite{hernandez_neutrino_2017}. The magnitude of the relative suppression also changes over time, leading to a redshift-dependence that also contributes to the uniqueness of the signal. 

The rest of this paper is organised as follows. Section \ref{section_methodology} provides a breakdown of our calculation method. Section \ref{section_results} provides our findings, with some details expanded upon in Section \ref{section_discussion}. Section \ref{section_conclusions} comprises our conclusions, and Appendix \ref{app_extended_results} summarises our results in table format.

\section{Methodology}\label{section_methodology}
\subsection{Overview}

\subsubsection{Model Parameters}

Our simplest fiducial model consists of the six standard $\Lambda$CDM parameters and an additional total neutrino mass parameter. Fiducial values for the $\Lambda$CDM parameters were extracted from \citep{planck_collaboration_planck_2016} based on the results from the \texttt{TT,TE,EE+lowP} dataset (see table 4 of \citep{planck_collaboration_planck_2016-1}). At certain points we expand this model to free the curvature parameter $\Omega_k$ and the dark energy equation of state parameter $w$ (which in some cases is allowed to be time-dependent). We assume a fiducial neutrino mass of 0.06 eV, which is close to the minimum limit implied by current neutrino oscillation experiments \cite[see, for example,][for a relatively recent review of neutrino mass hierarchy measurements]{qian_neutrino_2015}. We make the approximation of one massive neutrino and two massless neutrinos. Our Fisher matrix parameters and their fiducial values are summarised in Table \ref{table_fiducial_model}. Note that linear galaxy bias is also marginalised over as a nuisance parameter in all of our calculations, and its fiducial value varies depending on the survey and redshift bin (see details of surveys in Section \ref{section_results}).

\begin{table}[h!]
\centering
\begin{tabular}{l l r}
\hline
Parameter & Definition & Fiducial value\\
\hline
$\omega_b$ & Baryon density $\Omega_{b}h^2$ & 0.02225\\
$\omega_c$ & Cold dark matter density $\Omega_{c}h^2$ & 0.1198\\
100$\theta_{s}$ & $\theta_s$: Sound horizon size at last scattering (rad) & 1.04077\\
$\tau$ & Optical depth to last scattering & 0.079\\
$\ln(10^{10}A_s)$ & $A_s$: Amplitude of the primordial power spectrum & 3.094\\
$n_s$ & Spectral index of the primordial power spectrum & 0.9645\\
$M_\nu$(eV) & Total neutrino mass & 0.06\\
\hline
$w_{0}$ & Time-independent dark energy equation of state parameter & -1\\
$w_{a}$ & Time-dependent dark energy equation of state factor & 0\\
$\Omega_k$ & Curvature parameter & 0\\
\hline
\end{tabular}
\caption{Summary of the model parameters. The first seven parameters are always free, and the final three are free in some cases. We also marginalise over a free linear bias parameter in each redshift bin.}\label{table_fiducial_model}
\end{table}

\subsubsection{Priors}\label{subsection_method_priors}

All of our calculations are built upon a CMB prior. In the simplest case, we generate a minimalistic CMB \lq compressed likelihood\rq~prior from MCMC chains selected from the Planck Legacy Archive (the 
\noindent \texttt{base\_mnu\_plikHM\_TT\_lowTEB} dataset). The compressed likelihood prior compresses the information available from the CMB into four parameters that are effectively observables - the shift parameter $R = \sqrt{\Omega_{m}H_{0}^2}D_{A}(z_{*})/c$ (where $D_{A}$ is the angular diameter distance to the surface of last scattering), the angular scale of the sound horizon at last scattering $l_{A}=\pi/\theta_{s}$, $\omega_{b}$ and $n_{s}$ (see section 5.1.6 of \citep{planck_collaboration_planck_2016-1} for more information). We use the Fisher matrix mechanism to propagate these constraints into constraints on our cosmological parameter set. The advantage of the compressed likelihood prior is that the constraints it provides are relatively insensitive to variation in the curvature or dark energy equation of state. We also add very broad Gaussian priors on the parameters not constrained by the CMB prior to keep them within sensible ranges (see Section \ref{subsection_results_priors} for more information). In some cases, improvements on the constraints can be achieved by including information on parameters related to the amplitude of CMB fluctuations, $A_{s}\exp(-2\tau)$ and $\tau$, in the prior, and we highlight those cases.

\subsubsection{The Full Galaxy Power Spectrum Fisher Matrix}\label{sec_combined_constraints}

A Fisher matrix element for two parameters of the fiducial model indexed as $\alpha$ and $\beta$ can be calculated as

\begin{equation}\label{eq_fisher}
F_{\alpha\beta} = \pdv{\bar{\mathrm{O}}}{\theta_\alpha}\mathrm{C}^{-1}\pdv{\bar{\mathrm{O}}}{\theta_\beta}.
\end{equation}
Here $\bar{\mathrm{O}}$ is a vector of observable quantities, $\theta$ represents parameters of the fiducial model and $C$ is the covariance matrix of the observables.

To forecast the maximum amount of cosmological information available from a galaxy survey, constraints on the observed galaxy power spectrum $P_{g}(k,\mu)$ (where $\mu$ is the cosine of the angle with respect to the line of sight) are propagated directly into constraints on the cosmological parameters. The covariance of $P_{g}(k,\mu)$ can be most simply expressed for a given ($k, \mu$) increment as \cite[see, e.g.][]{percival_large_2013, font-ribera_desi_2014}

\begin{equation}\label{eq_covariance_matrix}
\langle \Delta P_{g}(k,\mu)^{2}\rangle = \frac{2\pi^2}{Vk^2\Delta k\Delta\mu}2P_{g}(k,\mu)^2,
\end{equation}
where $V$ is the volume of the redshift bin being observed, and $\Delta k$ and $\Delta \mu$ are the bin sizes for the wavenumber and angle with respect to the line of sight, respectively. Equation \ref{eq_covariance_matrix} applies in the case in which only one galaxy tracer population is assumed, with a single value assumed for the galaxy bias in each redshift bin. Here $P_{g}(k,\mu)$ is the full observed galaxy power spectrum including shot noise. Equation \ref{eq_covariance_matrix} can be appropriately generalised into a multi-dimensional band power matrix in cases in which multiple tracer populations (with different biases) are used, which also accounts for their cross-correlation. For a single galaxy tracer population, we calculate the galaxy power spectrum in a particular redshift bin (including linear RSD) and shot noise as

\begin{equation}\label{eq_full_pk}
P_g(k, \mu) = \left[ b+f(k)\mu^2\right]^{2}P_{\textrm{m}}(k) + \bar{n}_{g}^{-1}.
\end{equation}
Here $b$ is the fiducial bias of the galaxy sample, $f$ is the growth function with $f=\frac{\textrm{d}\ln D}{\textrm{d}\ln a}$ (where $D$ is the linear growth rate of perturbations) and $P_{\textrm{m}}$ is the real-space matter power spectrum. $\bar{n}_{g}$ is the galaxy number density and the final term accounts for shot noise. In the linear regime, $f$ is often taken as independent of scale, but massive neutrinos reduce the relative value of $f$ on small scales by a small amount, so we include this effect here.

To convert observational measurements into a galaxy clustering model, fiducial values of $H(z)$ and $D_{A}(z)$ must be assumed. If the product of $H(z)$ and $D_{A}(z)$ assumed is incorrect, the three-dimensional model will be distorted. This is the AP test, and it provides another source of constraints on our cosmological parameters. Therefore, as a final step, we convert our $k$ values into observable units and re-write the power spectra accordingly:

\begin{equation}
P(k_{\parallel}^{\textrm{obs}}, k_{\perp}^{\textrm{obs}}) =\frac{H(z)}{H_{fid}(z)}\left(\frac{D_{A, fid}(z)}{D_{A}(z)}\right)^2 P(k_{\parallel}^{\textrm{com}}, k_{\perp}^{\textrm{com}}),
\end{equation}
where $k_{\parallel}^{\textrm{obs}} = k_{\parallel}^{\textrm{com}}(H_{fid}(z)/H(z))$ and $k_{\perp}^{\textrm{obs}} = k_{\perp}^{\textrm{com}}(D_A(z)/D_{A,fid}(z))$.

The linear matter power spectra used in our calculations were all generated using \texttt{CLASS} \cite{blas_cosmic_2011}. To generate the fiducial $P_{\textrm{m}}(k)$ and $f(k)$ values as well as the numerical derivatives $\partial P_{\textrm{m}}/\partial\theta_{\alpha}$ and $\partial f/\partial\theta_{\alpha}$, we generated matter power spectra for a very dense sample of $z$ values, and stored the results in a two-dimensional table of $k$ and $z$ values. This table could then be interpolated to provide $P_{\textrm{m}}(z,k)$ values. Values of $D(k)$ could be extracted by dividing the power spectra, and $f(z,k) = \textrm{d}\ln{D(z,k)}/\textrm{d}\ln{a}$ could then be calculated.

Care was taken with derivatives to ensure that they were not very sensitive to the increments by which the parameters were varied in their calculation. Increments that are too small can result in numerical scattering, while those that are too large lose finer elements of the structure. Derivatives were generally calculated as $(P[\theta +\epsilon]-P[\theta-\epsilon])/2\epsilon$, with $P$ being either $P_{\textrm{m}}(k)$ or $f(k)$, and with $\epsilon$ taking the values outlined in Table \ref{table_derivative_increments}.

\begin{table}[h!]
\centering
\begin{tabular}{l l}
\hline
Parameter & Increment ($\epsilon$)\\
\hline
$\omega_b$ & 0.001\\
$\omega_c$ & 0.0025\\
100$\theta_{s}$ & 0.005\\
$\tau$ & 0.025\\
$\ln(10^{10}A_s)$ & 0.05\\
$n_s$ & 0.01\\
$M_\nu$(eV) & 0.02 \\
\hline
$w_{0}$ & 0.01\\
$w_{a}$ & 0.01\\
$\Omega_k$ & 0.01\\
\hline
\end{tabular}
\caption{List of the increment sizes used to calculate the derivatives numerically for each parameter.}\label{table_derivative_increments}
\end{table}

All of our constraints in this paper are calculated with marginalisation over the linear galaxy bias $b$. Derivatives with respect to the bias parameter(s) can be calculated analytically using Equation \ref{eq_full_pk}. For a single tracer population:

\begin{equation}\label{eq_derivative_bias}
\pdv{P_{g}(k, \mu)}{b} = 2\left[b+f(k)\mu^2\right]P_{\textrm{m}}(k).
\end{equation}

It is important to define maximum and minimum usable $k$ values in each redshift bin of a survey. $k_{\textrm{min}}$ is calculated based on the dimensions of a particular redshift bin. $k_{\textrm{max}}$ is a scale beyond which non-linear effects are too strong for linear approximations to be accurate. We choose $k_{\textrm{max}}$ = 0.2 $h$ Mpc$^{-1}$ here. In the case of BAO-only projections (the BAO signal is particularly robust against non-linear effects), it is common practice to replace the sharp $k$-cutoff with an exponential degradation factor in the signal, which replicates a gradual smearing effect on the BAO peaks (see \cite{eisenstein_robustness_2007, seo_improved_2007}). In \cite{eisenstein_improving_2007}, it was suggested that the degraded BAO information could be recovered by reconstructing the original linear density field for a particular galaxy survey by using knowledge gleaned from the galaxy distribution to reverse the displacements of galaxies due to bulk flows and cluster formation. In our BAO-only calculations, we replace the sharp $k$ cut-off with an exponential damping factor given by

\begin{equation}\label{eq_exp_damping}
P_{\textrm{BAO, damped}}(k, \mu) = P_{\textrm{BAO}}\exp\left[-\frac{1}{2}\left(k_{\parallel}^2\Sigma_{\parallel}^2+k_{\perp}^2\Sigma_{\perp}^2\right)\right].
\end{equation} 

The damping scales $\Sigma_{\parallel}$ and $\Sigma_{\perp}$ are calculated as a function of the structure growth rate $f(z)$ and the amplitude of the power spectrum $\sigma_{8}(z)$ as described by \cite{seo_improved_2007}. We follow the example of \cite{font-ribera_desi_2014} to account for the possibility of improving constraints with reconstruction. We multiply the damping scales for a given redshift bin by a reconstruction factor $r$ calculated using the value of $n_{g}P_g(k=0.14~h~\textrm{Mpc}^{-1},~\mu=0.6)$ in that bin. For high-density bins, $r$ reaches a maximum of 0.5, while in low-density bins it is just 1. For intermediate values, we interpolate over the same table of values given by \cite{font-ribera_desi_2014}. In all other cases (beyond BAO), we use a sharp cut-off at 0.2 $h$ Mpc$^{-1}$.

\subsubsection{Removing Baryonic Oscillations from the Matter Power Spectrum}\label{sec_smoothing}

In the following sections, we attempt to determine the constraints that can be placed on the sum of the neutrino masses using different elements of the observed galaxy power spectrum. The sinusoidal BAO signal varies in both its phase and amplitude with many of our cosmological parameters. In some cases we need to remove the BAO signal from our Fisher derivatives of the matter power spectrum to isolate other effects, or to isolate the effects on the BAO signal alone. 

We can consider the matter power spectrum to consist of two components, a BAO component and a smooth component (S): $P_{\textrm{m}}(k) = P_{\textrm{S}}(k) + P_{\textrm{BAO}}(k)$. There are several common methods of extracting $P_{\textrm{S}}$ (which can then be subtracted to obtain $P_{\textrm{BAO}}$ alone), including fitting a spline to $P_{\textrm{m}}(k)$ that passes through the zero-points of the BAO oscillation, or using a formula for calculating the smooth power spectrum like that provided by \cite{eisenstein_baryonic_1998}. These methods are unsuitable in our case, however, as we require the derivatives of $P_{\textrm{BAO}}$ or $P_{\textrm{S}}$ for insertion into the Fisher matrix, and small inaccuracies in the fitting of the matter power spectra can lead to artificially large or distorted derivatives. Therefore, we first calculate the derivatives of the full $P_{\textrm{m}}(k)$, and then apply a smoothing function to the derivatives themselves to extract the smooth part of the derivative, which can be subtracted from the full derivative to obtain the derivative of the oscaillatory part.

As a smoothing function we use a Savitzky-Golay filter \cite{savitzky_smoothing_1964}. The Savitzky-Golay filter sees the BAOs as noise and because of its averaging technique provides more consistent results than spline-fitting, which depends on manual selection of zero-point $k$ values by sight. This application of the Savitzky-Golay method can be validated by applying it to a fiducial power spectrum (rather than a derivative) and then subtracting the fit from the original data to show a very regular and smooth BAO signal. In the case of derivative fitting, the smoothing is done in $\textrm{d}\ln P(k)/\textrm{d}\theta$ - log $k$ space, and then both the original and smoothed spectra are plotted with a linear $P(k)$ scale to ensure that the fit remains reasonable. The derivative of the BAO component can then be obtained via subtraction and inspected. 

\subsection{Distance Information}\label{section_methodology_distance}

The most popular distance probe used to constrain cosmological parameters is the BAO signature. However, the broadband galaxy power spectrum also provides other means of constraining the cosmological distance parameters $H(z)$ and $D_{A}(z)$ \cite{shoji_extracting_2009}. The AP test requires $H(z)$ and $D_{A}(z)$ to scale appropriately with each other so that cosmological isotropy is preserved in real space. Other characteristic scales in the matter power spectrum, including the matter-radiation equality peak and the Silk damping scale, also provide distance constraints. 

\subsubsection{Full Distance Constraints}
To extract the maximum amount of distance information from the galaxy power spectrum, we first use Equation \ref{eq_covariance_matrix} to extract constraints on $P_{g}(k, \mu)$. We then propagate these constraints into constraints on $\ln H(z)$ and $\ln D_{A}(z)$ (marginalised over bias $b$, the growth factor $f$ and the underlying matter power spectrum $P_{\textrm{m}}$) using the following derivatives (see e.g. \cite{shoji_extracting_2009}):

\begin{equation}\label{eq_der_da}
\pdv{P_{g}(k,\mu)}{\ln D_{A}} = \pdv{P_{g}(k,\mu)}{k}\pdv{k}{\ln D_{A}} +  \pdv{P_{g}(k,\mu)}{\mu^2}\pdv{\mu^2}{\ln D_{A}},
\end{equation}
\begin{equation}\label{eq_der_h}
\pdv{P_{g}(k,\mu)}{\ln H} = \pdv{P_{g}(k,\mu)}{k}\pdv{k}{\ln H} +  \pdv{P_{g}(k,\mu)}{\mu^2}\pdv{\mu^2}{\ln H}.
\end{equation}

The derivatives of $P_{g}(k,\mu)$ with respect to $k$ and $\mu^2$ can be obtained directly from the calculated fiducial $P_{g}(k, \mu)$. The other terms are easily derived analytically:

\begin{equation}
\pdv{k}{\ln D_{A}} = k(1-\mu^2);~\pdv{k}{\ln H} = -k(\mu^2);~\pdv{\mu^2}{\ln D_{A}}=\pdv{\mu^2}{\ln H} = -2\mu^2(1-\mu^2).
\end{equation}

\subsubsection{BAOs}
In the case that we want the information from the BAO signal alone, we must apply the method outlined in Section \ref{sec_smoothing} to replace the full derivative of the matter power spectrum with just the oscillatory part when calculating the derivatives of $P_{g}(k,\mu)$ as above.

Seo and Eisenstein \cite{seo_improved_2007} provided a useful fitting function for forecasting $H$ and $D_{A}$ constraints from the BAO signal alone. We use our own fitting method here because it was most compatible in the context of our code, but our results agree well with published forecasts that use the Seo and Eisenstein method to predict constraints on $H(z)$ and $D_{A}(z)$, including a consistent $\ln H$-$\ln D_{A}$ correlation factor of 0.4. For our BAO-only constraints, we remove the limit on $k_{\textrm{max}}$ and instead enforce an exponential decay factor as described in Section \ref{sec_combined_constraints}. 

\subsubsection{AP Test}
The AP test provides constraints on $H(z)$ and $D_{A}(z)$ by requiring that these values scale appropriately to preserve isotropy when the observed galaxy power spectrum is converted into real space coordinates. If the assumed product $H(z)D_{A}(z)$ is wrong, anisotropies will appear in the model. For AP information to be at its strongest, the redshift-space distortion effect must be well constrained so the two effects can be distinguished. Here, we extract tightest constraints that would be available from the AP test alone by holding the redshift space distortions fixed. 

The AP test provides its constraints through changes in the observed galaxy power spectrum with the observation angle. The derivatives used are therefore as follows: 
 
\begin{equation}
\pdv{P_{g}(k,\mu)}{\ln D_{A}} = \pdv{P_{g}(k,\mu)}{\mu^2}\pdv{\mu^2}{\ln D_{A}},
\end{equation}
\begin{equation}
\pdv{P_{g}(k,\mu)}{\ln H} = \pdv{P_{g}(k,\mu)}{\mu^2}\pdv{\mu^2}{\ln H}.
\end{equation}

When the AP test is used alone, it provides a correlation coefficient between $H$ and $D_{A}$ of $-1$ \cite{shoji_extracting_2009}. Adding further distance information, such as standard rulers in the shape of the matter power spectrum, constrains $H$ and $D_{A}$ individually and allows this degeneracy to be broken. 

\subsection{Structure Growth}

\subsubsection{RSD}

The RSD is an anisotropy that arises in the observed redshift-space power spectrum because the measured redshift of a particular galaxy is a function not just of the Hubble flow but also its peculiar velocity. The $\left[ b+f(k)\mu^{2}\right]$ factor of Equation \ref{eq_full_pk} is used to account for the Kaiser effect \cite{kaiser_clustering_1987} resulting from structure formation, which causes an apparent strengthening of the clustering amplitude along the line of sight as objects fall into high-density regions. Galaxy survey measurements can be used to constrain $f\sigma_{8}$, where $f$ is the structure growth rate and $\sigma_{8}$ the normalisation of the power spectrum amplitude, through analysis of this anisotropic signal.

White \textit{et al.} \cite{white_forecasting_2009} previously provided a method for isolating the information available from RSD in galaxy surveys, isolating $f$ and keeping $\sigma_{8}$ fixed. We follow their example here, but include the scale-dependence of $f(k)$, and marginalise over both the bias and the entire matter power spectrum (which includes $\sigma_{8}$), using the following derivatives:

\begin{equation}
\pdv{P_{g}(k,\mu)}{\theta} = \pdv{P_{g}(k,\mu)}{f(k)}\pdv{f(k)}{\theta} = \left[\frac{2\mu^2}{b+f(k)\mu^2}\right]P_g(k,\mu)\pdv{f(k)}{\theta},
\end{equation}

\begin{equation}
\pdv{P_{g}(k,\mu)}{P_{\textrm{m}}(k)} = \left[b+f(k)\mu^2\right]^2.
\end{equation}

We also marginalise over the distance parameters $H(z)$ and $D_{A}(z)$ using the derivatives given in Equations \ref{eq_der_da} and \ref{eq_der_h}. We can also choose to extract constraints using the product $f\sigma_{8}$ if we do not wish to isolate the effect on the structure growth rate. $\sigma_{8}$ is calculated as an integral over the matter power spectrum and therefore also theoretically contains information on the suppression of the small-scale matter power by massive neutrinos. 

\subsubsection{The Small-Scale Suppression of the Structure Growth Rate}

\begin{figure}
\includegraphics[width=\textwidth]{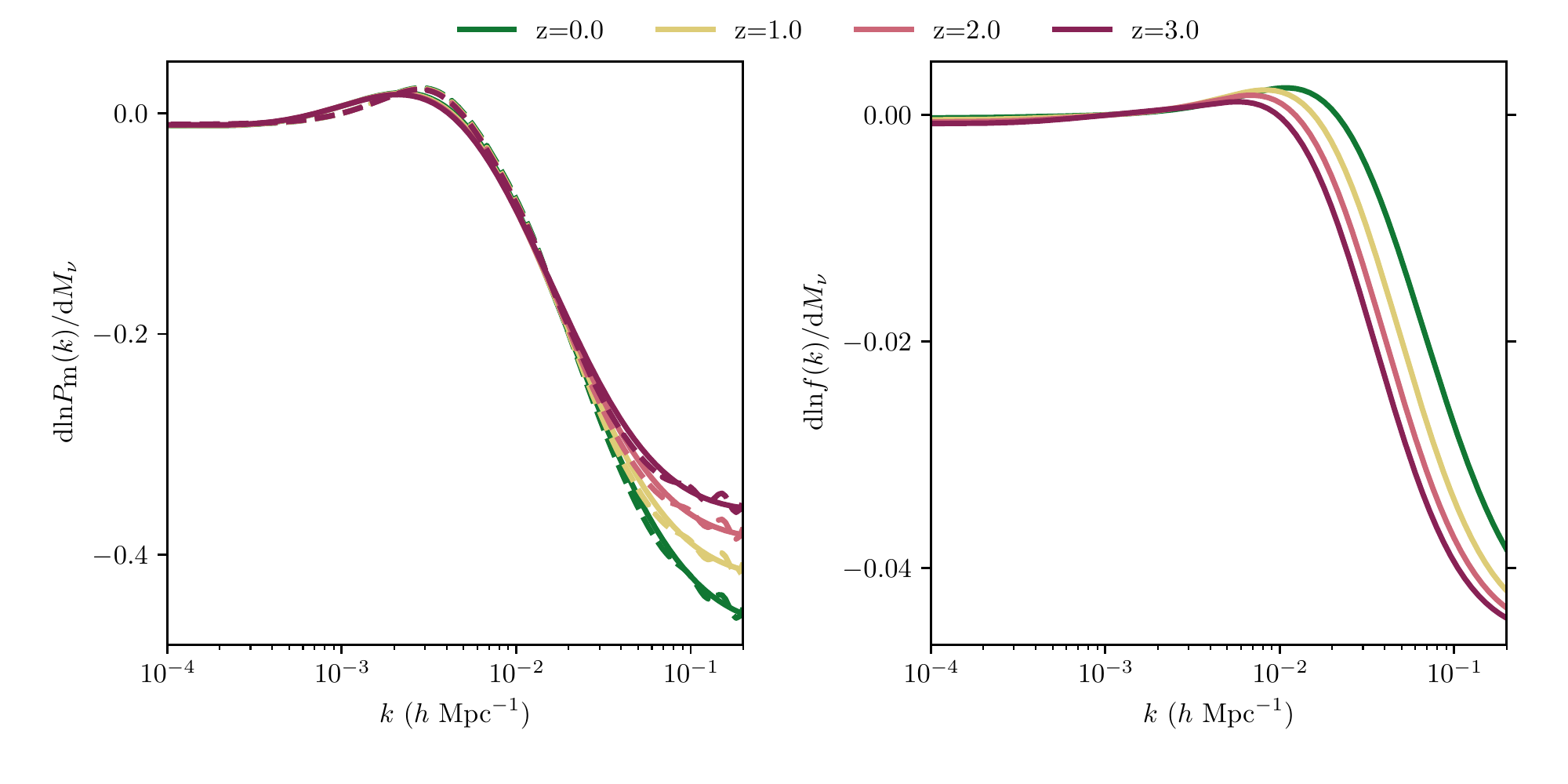}
\caption{The effect of increasing neutrino mass on the shape of the matter power spectrum $P_{\textrm{m}}$ (left) and on the structure growth rate $f$ (right), both normalised so that the amplitude of $P_{\textrm{m}}(k)$ and $f(k)$ at $k=10^{-3}$ $h$ Mpc$^{-1}$ is fixed. The magnitude of the relative suppression increases with time in the case of the matter power spectrum because the effects of neutrino free-streaming have had longer to accumulate. On the other hand, the relative suppression decreases with time in the structure growth rate case as the neutrinos become non-relativistic. The matter power spectrum derivative has been smoothed to remove the BAO signal, and the dashed lines underneath show the unsmoothed case.}\label{fig_derivatives}
\end{figure}

The treatment described in the previous section uses information from both the (constant) large-scale value of $f(k)$ and the small-scale suppression of $f(k)$ to derive constraints. We can now isolate the information available from the scale-dependent component of $f(k)$ alone. We re-write $f(k)$ as 

\begin{equation}
f(k) = f_1.f_2(k).
\end{equation}

Here $f_1$ is the value of $f$ on large scales ($k=10^{-3}$ $h$ Mpc$^{-1}$ to be specific, where it is still constant, although the exact scale chosen should not matter) and $f_{2}(k)$ is a scale-dependent correction factor (equal to one on large scales). To isolate the information from the neutrino suppression of $f$ on small scales, we replace $f$ in our derivatives with $f_{2}$ multiplied by the fiducial value of $f_{1}$ and marginalise over $f_{1}$ as an additional Fisher matrix parameter alongside the other cosmological parameters. In other words, $f_{1}$ is taken as a constant outside the derivative and $df/d\theta_{\alpha}$ becomes $f_{1}.(df_{2}/d\theta_{\alpha})$. This method removes any assumptions about the large-scale value of $f$ and therefore can allow for the possibility of alternative gravity models. The right panel of Figure \ref{fig_derivatives} shows the derivatives of $f(k)$ with respect to the neutrino mass with $f_{1}$ held constant at a range of redshifts.

\subsubsection{The Small-Scale Suppression of the Matter Power Spectrum}\label{subsec_shape}

In order to obtain the information contained in the power spectrum shape alone, we must exclude the information from the BAO feature, which we have already accounted for. We extract the derivatives of $P_{\textrm{s}}$ with respect to the cosmological parameters following the method outlined in Section \ref{sec_smoothing}. In the rest of this section, $P_{\textrm{m}}$ refers to the smoothed matter power spectrum. 

As in the case of the structure growth factor $f$, we want to extract the scale-dependent component of $P_{\textrm{m}}$ alone as an information source, neglecting the amplitude of the power spectrum (see Figure \ref{fig_derivatives}). As before, we introduce a new parameterisation

\begin{equation}
P_{\textrm{m}}(k) = P_{1}.P_{2}(k),
\end{equation}
where $P_{1}$ is the matter power spectrum value at  $k=10^{-3}$ $h$ Mpc$^{-1}$ (the point of normalisation should not matter, and we try to avoid the edges where interpolation effects are stronger). The derivative of our Fisher matrix becomes:

\begin{align}
\pdv{P(k, \mu)}{\theta_{\alpha}} = \pdv{P(k, \mu)}{P_{\textrm{m}}}\pdv{P_{\textrm{m}}}{\theta_{\alpha}} = \left[b+f(k)\mu^2\right]^{2}P_1\pdv{P_{2}}{d\theta_{\alpha}}
\end{align}

The derivatives of $P_{\textrm{m}}$ with respect to $M_{\nu}$ obtained from \texttt{CLASS} are provided in Figure \ref{fig_derivatives} with $P_1$ held fixed. Both the unsmoothed and smoothed fits are shown. We marginalise over $P_{1}$, bias, the structure growth rate $f$, and the distance parameters $H(z)$ and $D_{A}(z)$.

\section{Results}\label{section_results}

\subsection{Survey Data}

In this section, we reference several current and upcoming galaxy surveys. We briefly describe the survey parameters assumed in this sub-section.

Our \textbf{HETDEX} \cite{hill_hobby-eberly_2008} survey model consists of two redshift bins, with a total redshift range of $1.9\leq z\leq 3.5$ and an area of 425 deg$^2$. The model comprises a total volume of 2 Gpc$^3/h^3$ and a total of 0.8 million galaxies \cite{leung_bayesian_2017}. We assume a constant bias of $b(z) = 1.5$.

For \textbf{PFS}, we use the survey parameters and bias values specified in table 2 of \cite{takada_extragalactic_2014}. Under these specifications, PFS will have a redshift range of $0.6\leq z\leq 2.4$ and an area of 1464 deg$^2$, providing a total volume of 9.91 Gpc$^3/h^3$ and 4.18 million galaxies across all redshift bins.

For \textbf{DESI}, we refer to table 2.3 of \cite{desi_collaboration_desi_2016}. The authors provide a range of survey plans. The main survey covers a redshift range of $0.6\leq z\leq 1.9$ with an area of 14000 deg$^2$. Redshift bin volumes are provided in units of Gpc$^3/h$ so we recalculated them in units of Gpc$^3/h^3$ to comply with our code. The total volume was then calculated to be 57.36 Gpc$^3/h^3$ with 22.35 million galaxies in total for three individual galaxy populations - emission line galaxies (ELGs), luminous red galaxies (LRGs) and quasars (QSOs). In some cases we use data from only the ELGs for a more direct comparison with other surveys. We use the fiducial bias formulae provided in section 2.4.2 of \cite{desi_collaboration_desi_2016}.

There is much less specific survey information available for \textbf{Euclid} \cite{amendola_cosmology_2016, laureijs_euclid_2011} and \textbf{WFIRST} \cite{spergel_wide-field_2013, green_wide-field_2012}. We follow tables 6 and 7 of \cite{font-ribera_desi_2014}, respectively, for these two surveys. In the case of Euclid, this assumes a survey area of 15000 deg$^2$ and a redshift range of $0.6\leq z\leq 2.1$, corresponding to a survey volume of 72 Gpc$^3/h^3$ with a total galaxy count of 50 million galaxies. We assume a bias scaling of $b(z)D(z) = 0.76$. For WFIRST, the survey area is 2000 deg$^2$ and the redshift range is $1\leq z\leq 2.8$. The survey volume is 13.55 Gpc$^3/h^3$, containing 26.5 million galaxies. We calculate the bias in this case as $b(z) = 1.5 + 0.4(z-1.5)$.

In each of the following subsections, we analyse in depth our forecasts for Euclid, the strongest of the surveys we consider. We also provide a summary of the constraints for the other surveys outlined here. 

\subsection{Priors}\label{subsection_results_priors}

We should begin by understanding how parameters are correlated with each other in the CMB prior. The prior only really constrains $\theta_{s}$, $n_{s}$ and $\omega_{b}$. In some cases, we demonstrate the effect of adding information on parameters related to the amplitude of the fluctuations of the CMB: $A_{s}\exp(-2\tau)$ and $\tau$ (see Sections \ref{subsubsection_rsd} and \ref{subsection_priors}). The remaining parameters - $M_{\nu}$, $\omega_{cdm}$, $\Omega_{k}$, $w_{0}$ and $w_{a}$ - all modify $R$, but as the other prior parameters are held fixed, their effects are completely degenerate and they are unconstrained from the prior alone without additional information. A useful study of the degeneracies between $M_\nu$ and parameters in CMB data is provided in \cite{archidiacono_physical_2017}. We add wide Gaussian priors on these unconstrained parameters to keep their values sensible. The general intention is that the results that follow should be relatively independent of the exact priors chosen. Table \ref{Table_priors} summarises these prior values.

\begin{table}
\centering
\begin{tabular}{c | c}
\hline
Parameter & 1-$\sigma$ error\\
\hline
$\ln(10^{10}A_{s})$ & 1.0\\
$\tau$ & 0.5\\
$M_{\nu}$ (eV) & 1.0\\
$\omega_{cdm}$ & 0.2\\
$\Omega_{k}$ & 0.1\\
$w_{0}$ & 1.0\\
$w_{a}$ & 3.0\\
\hline
\end{tabular}
\caption{1-$\sigma$ uncertainties imposed on the cosmological parameters as initial priors. $\theta_{s}$, $n_{s}$ and $\omega_{b}$ are constrained by the CMB prior.}\label{Table_priors}
\end{table}

\subsection{Distance Information: BAO and AP}

Constraints on the distance parameters $H(z)$ and $D_{A}(z)$ are derived from two main sources in galaxy surveys. The AP test constrains the product of $H(z)$ and $D_{A}(z)$ by requiring that the galaxy clustering pattern derived from observations be isotropic. The BAO scale imprinted on the galaxy clustering pattern breaks the degeneracy between $H$ and $D_{A}$ and allows them to be measured individually. Roughly speaking, the AP test constrains $D_{A}(z)H(z)$ and the BAO signal constraints $D_{A}(z)^2/H(z)$ \cite[e.g.][]{shoji_extracting_2009}. 

\begin{figure}
\includegraphics[width=\textwidth]{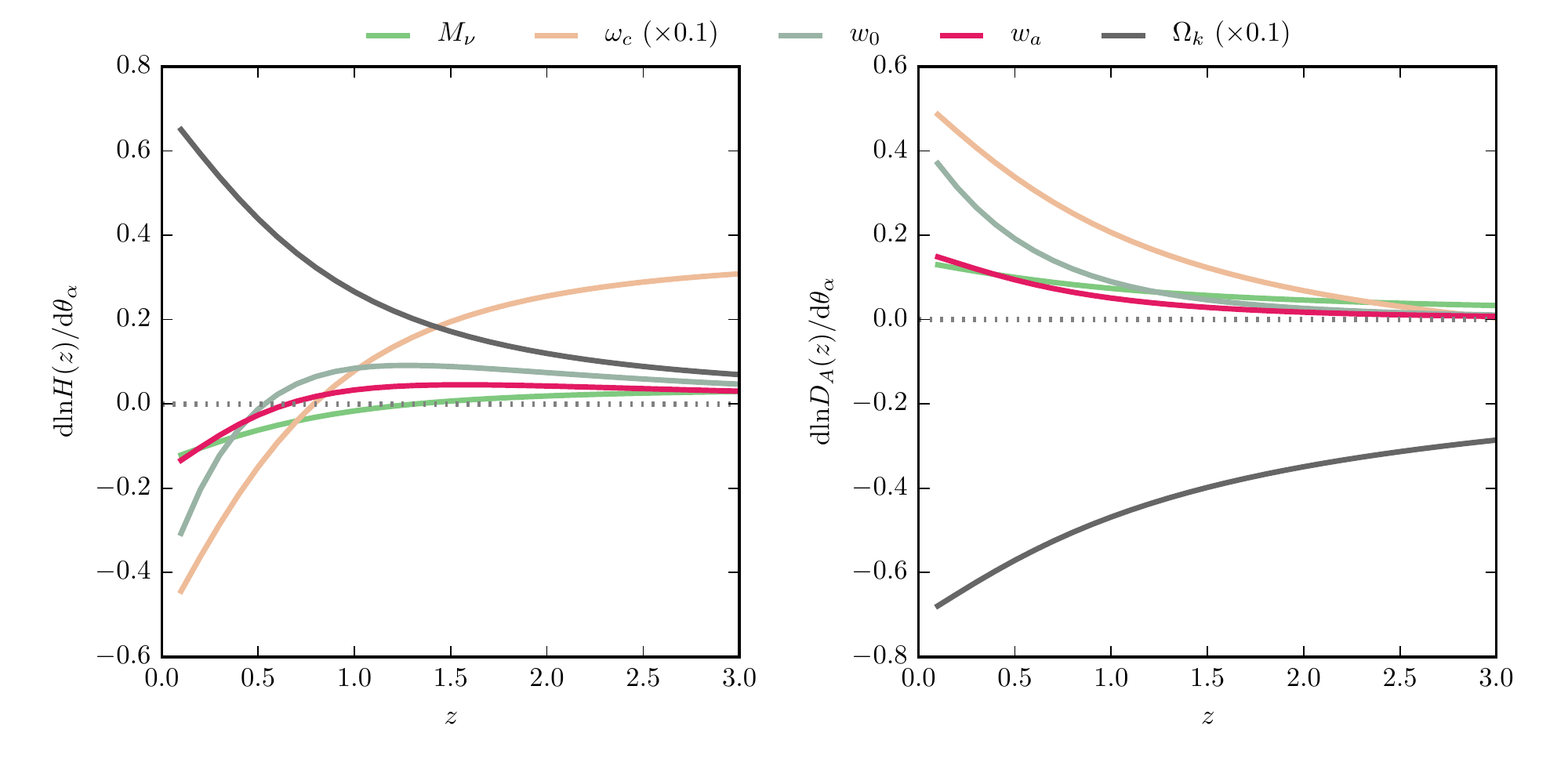}
\caption{The derivatives of $H(z)$ and $D_{A}(z)$ with respect to the parameters $M_{\nu}$, $\omega_{\textrm{cdm}}$, $w_{0}$, $w_{a}$ and $\Omega_{k}$. The effects of adding extensions to the $\Lambda$CDM model on the neutrino mass constraints can be understood by comparing these derivatives. The derivatives with respect to $\omega_{cdm}$ and $\Omega_{k}$ have been re-scaled by a factor of 0.1 for plotting purposes.} \label{fig_derivatives_distance}
\end{figure}

$H$ and $D_{A}$ change with $\theta_{s}$ and the matter density. The strength of distance information in constraining the neutrino mass lies in its ability to break the correlation between $\omega_{cdm}$ and $M_{\nu}$ in the CMB prior. The effects of increasing either $\omega_{cdm}$ or $M_{\nu}$ on $H$ and $D_{A}$ are strongly degenerate, leading to strong anti-correlation between the two parameters that breaks the degeneracy created by the CMB constraints. The associated disadvantage, however, is that distance information is in fact sensitive to the sum of $M_{\nu}$ and $\omega_{cdm}$, of which $M_{\nu}$ makes up a tiny fraction. Understanding the relationship between $M_{\nu}$ and $\omega_{cdm}$ is the key to understanding the constraints on $M_{\nu}$ provided by distance probes. Figure \ref{fig_derivatives_distance} shows the effect of changing these parameters on $H$ and $D_{A}$ as a function of redshift.

Increasing $\omega_{cdm}$ or $M_{\nu}$ in the $\Lambda$CDM context requires that $\Omega_{\Lambda}$ is decreased to maintain the critical energy density. At higher redshifts, the increase in $\Omega_{m}$ is the dominant effect on $H(z)$, which is then increased relative to the fiducial model. In the later, dark-energy-dominated regime, the decrease in $\Omega_{\Lambda}$ dominates the change in $H(z)$, which is now reduced compared to in the fiducial model. The effects are not completely degenerate as the crossover occurs at a higher redshift with additional massive neutrinos than with additional cold dark matter, as a result of the historical relativistic nature of the massive neutrinos. These changes in $H(z)$ mean that $D_{A}(z)$ is increased relative to the fiducial model at the redshifts covered by our surveys. The similarity in the effects of increasing $\omega_{cdm}$ and $M_{\nu}$ on the distance parameters (i.e. both increase $H$ at high $z$ and decrease it at low $z$, and both increase $D_{A}$, particularly at late times) makes the two parameters highly anti-correlated. 

Figure \ref{fig_derivatives_distance} also shows the impact of varying $w_{0}$, $w_{a}$ and $\Omega_{k}$ on the distance parameters, helping us understand how freeing these parameters can affect the constraint on $M_{\nu}$. Increasing $w_{0}$ or $w_{a}$ reduces $H(z)$ in the later, dark-energy-dominated regime, but increases it at higher redshifts. This effect is qualitatively similar to that of increasing $M_{\nu}$ and $\omega_{cdm}$, and the corresponding effects on $D_{A}(z)$ are also similar. Freeing $w$ therefore weakens the constraint on $\omega_{cdm}$ considerably. However, the constraints on $M_{\nu}$ are much less affected because $M_{\nu}$ starts to suppress $H(z)$ at much higher redshifts, and the slope of the derivative is much less steep than for the other parameters, allowing this effect to be distinguished. 

On the other hand, Figure \ref{fig_derivatives_distance} shows that the shape of the derivative of $\Omega_{k}$ is very similar to that of $M_{\nu}$, though inverted. The effects of $\Omega_{k}$ and $M_{\nu}$ on both distance parameters are quite clearly degenerate, and the effect of an increase in $\Omega_{k}$ could quite clearly be compensated by a reduction in $M_{\nu}$, and vice versa. These effects are reflected in the results we obtain. 

\begin{figure}
\includegraphics[width=\textwidth]{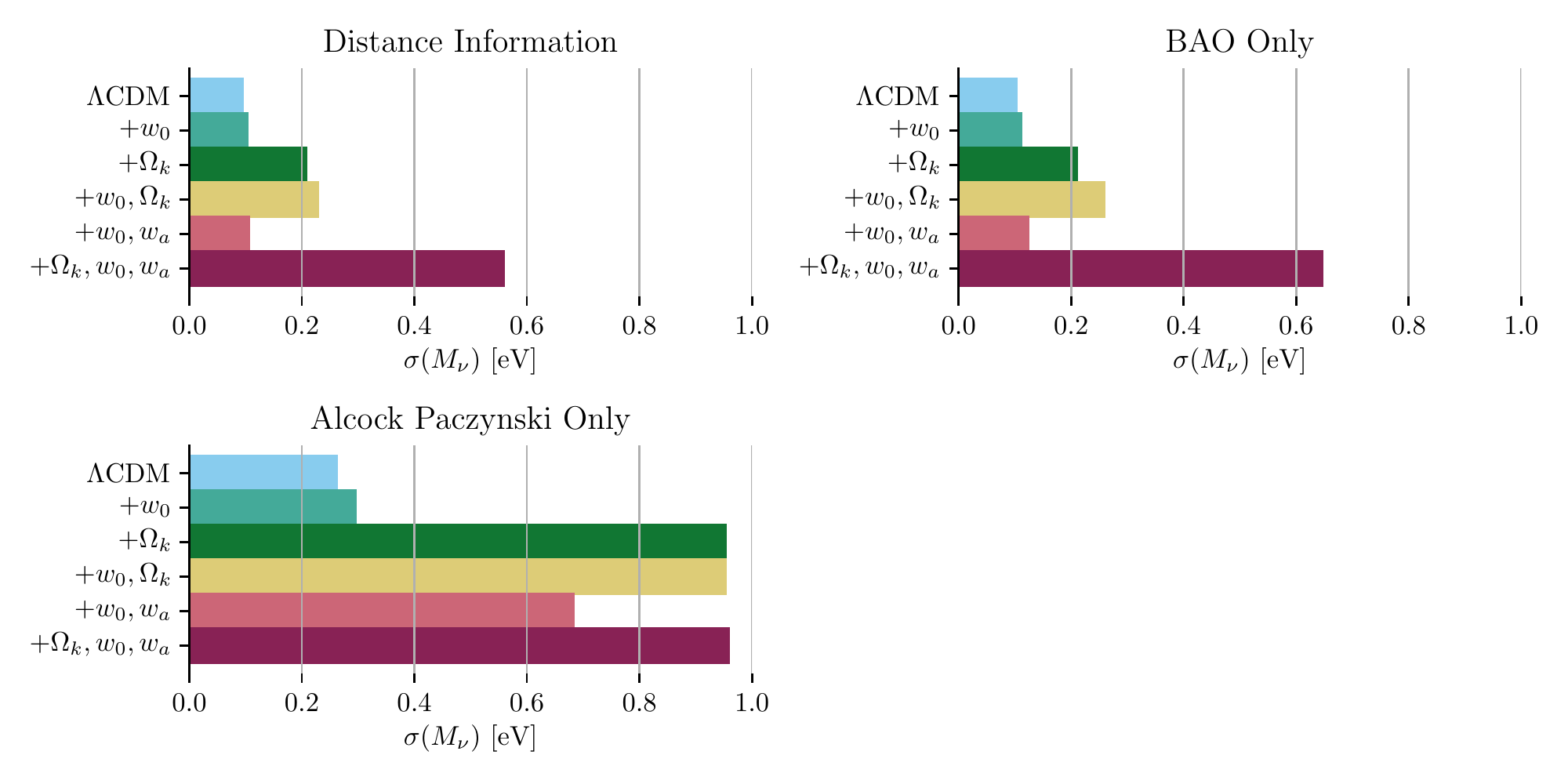}
\caption{A breakdown of the forecasted distance information constraints on $M_{\nu}$ for Euclid (including a CMB prior on $\theta_{s}$, $n_{s}$ and $\omega_{b}$), for a variety of models. It is clear that the primary source of constraining information is the BAO signal. All of the constraints are weakened considerably if $\Omega_{k}$ is allowed to vary, and in the AP only case this results in effectively no constraint.}\label{fig_distance}
\end{figure}

Figure \ref{fig_distance} shows a breakdown of the distance constraints on $M_{\nu}$ for a series of cosmological models. It is clear that the vast majority of the distance information comes from the BAO signal. It is also clear that in all cases, the constraints are significantly degraded when $\Omega_{k}$ is allowed to vary. If BAO information is included, the constraints are not very sensitive to assumptions about the dark energy equation of state.

\begin{figure}
\includegraphics[width=\textwidth]{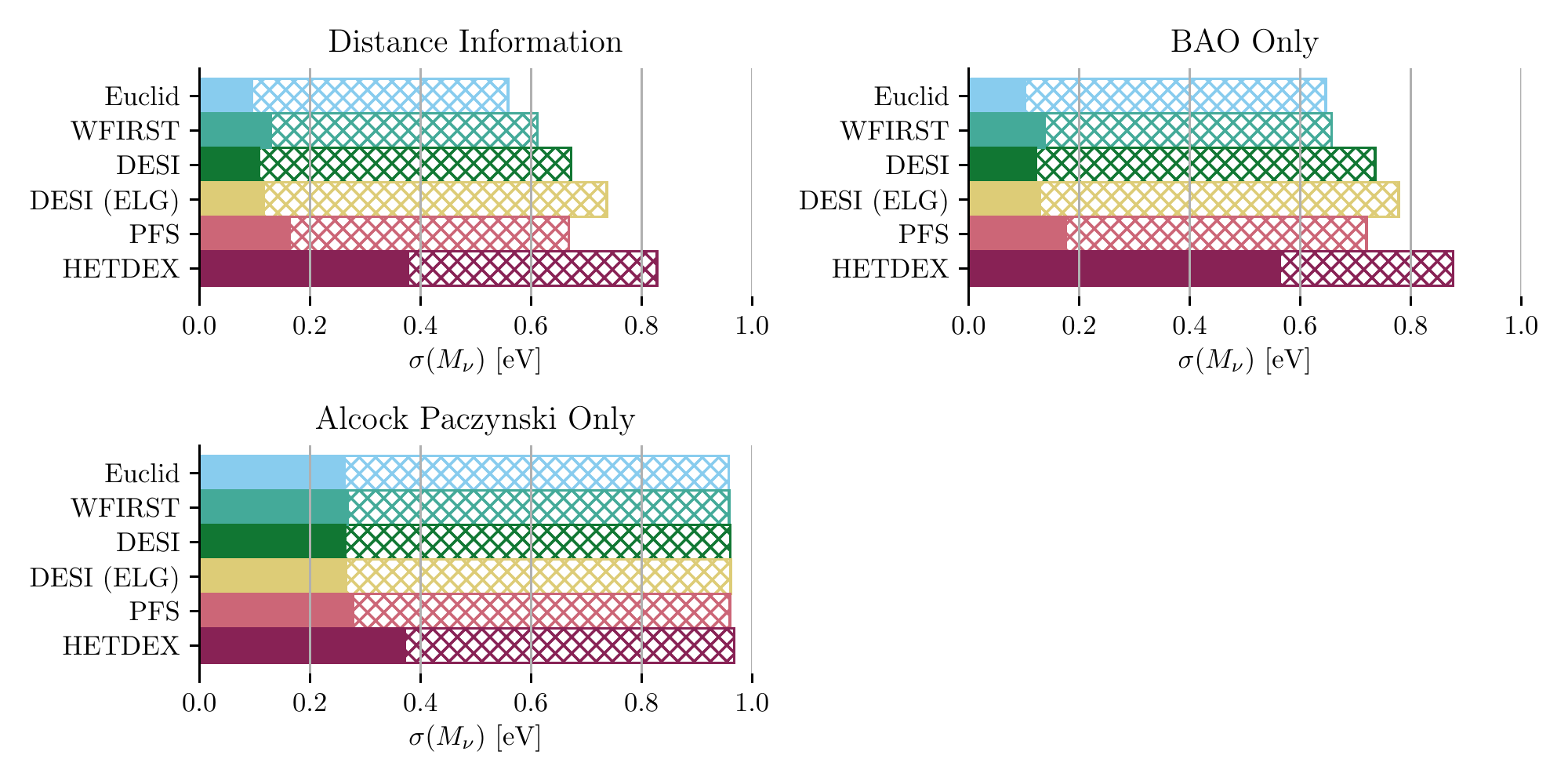}
\caption{The same breakdown as in Figure \ref{fig_distance}, but this time for all surveys. The constraints are shown here for two cosmological models - solid bars represent the $\Lambda$CDM model and the cross-hatched bars give the constraints for the most complex model (+$\Omega_{k}$, $w_{0}$, $w_{a}$). Note that the DESI survey plan makes use of three galaxy tracer populations, and DESI (ELG) refers to the case in which only the Emission Line Galaxies (ELG) are taken into account.} \label{fig_distance_surveys}
\end{figure}

Figure \ref{fig_distance_surveys} shows the distance constraints for different surveys. The solid bars represent the $\Lambda$CDM constraint, and the cross-hatched bars the most complicated model (+$\Omega_{k}$, $w_{0}$, $w_{a}$). It is clear that the constraints are heavily model-dependent in all cases. 

With the BAO feature smoothed out, $H(z)$ and $D_{A}(z)$ become strongly anti-correlated, due to the AP test dominating the remaining information. Using the BAO feature alone, there is a consistent correlation coefficient of approximately 0.4. With all of the distance information, the correlation is approximately $-0.4$ to $-0.5$.

\subsection{Structure Growth Information}

Sources of information on the growth of structure include the shape and amplitude of the matter power spectrum, and the observed anisotropies created by RSD.

\subsubsection{RSD}\label{subsubsection_rsd}

We can use RSD to extract constraints on the product $f\sigma_{8}$ or on $f$ marginalised over $\sigma_{8}$ if we want to isolate constraints from the structure growth rate alone. It is important to note that $\sigma_{8}$ is calculated as an integral over the matter power spectrum, and therefore also contains information on the overall shape of the matter power spectrum, which is altered by massive neutrinos. The relationship between $\sigma_{8}$ and $M_{\nu}$ is therefore quite complex.

\begin{figure}
\includegraphics[width=\textwidth]{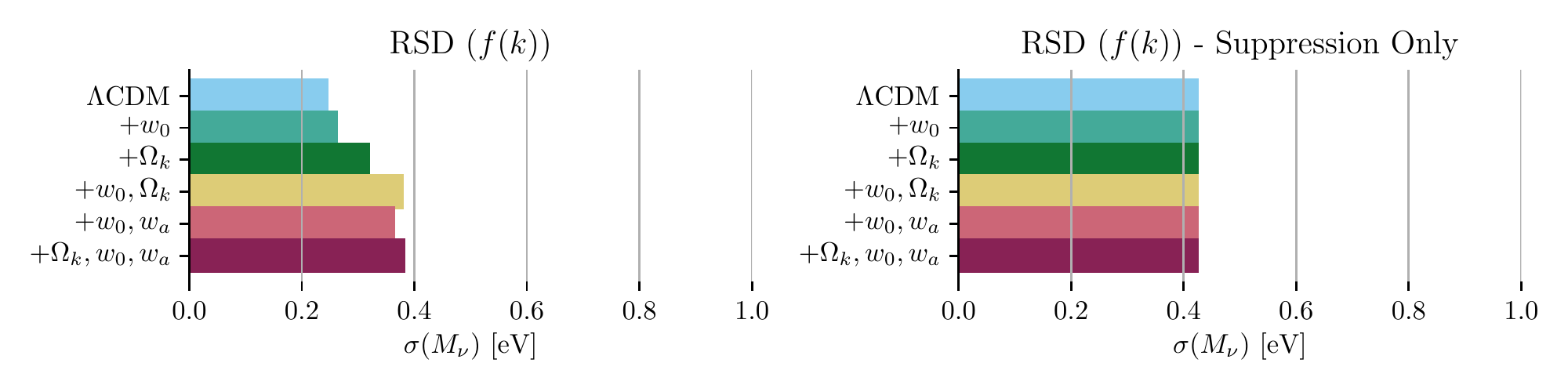}
\caption{A breakdown of the forecasted constraints on $M_{\nu}$ from RSD for Euclid (including a CMB prior on $\theta_{s}$, $n_{s}$ and $\omega_{b}$), for a variety of models. In both cases, RSD are used to constrain the structure growth rate $f$ marginalised over the matter power spectrum (which includes $\sigma_{8}$). The right-hand panel gives the constraints from the scale-dependence of $f(k)$ alone (the large-scale amplitude of $f(k)$ is taken outside the derivatives and marginalised over). The left-hand panel gives the constraints available from both the large-scale amplitude and scale-dependence of $f(k)$.}\label{fig_f}
\end{figure}

We begin by considering the information available from constraints on $f$ alone, marginalised over $\sigma_{8}$. Figure \ref{fig_f} provides the key results for Euclid in two different cases. The left-hand panel demonstrates the constraining power of $f(k)$ values at all scales (including both the large-scale values and the relative change on small scales characteristic of massive neutrinos). We see that the constraints vary significantly depending on the cosmological model adopted. The right-hand panel shows the constraints if only the scale-dependence of $f$ is considered, with the large-scale value of $f$ being marginalised over. Here we see that the results are completely independent of the choice of cosmological model. This is because the scale-dependence of the structure-growth rate probed is a unique indicator of massive neutrinos, and is not replicated by the additional parameters we can include. 

\begin{figure}
\includegraphics[width=\textwidth]{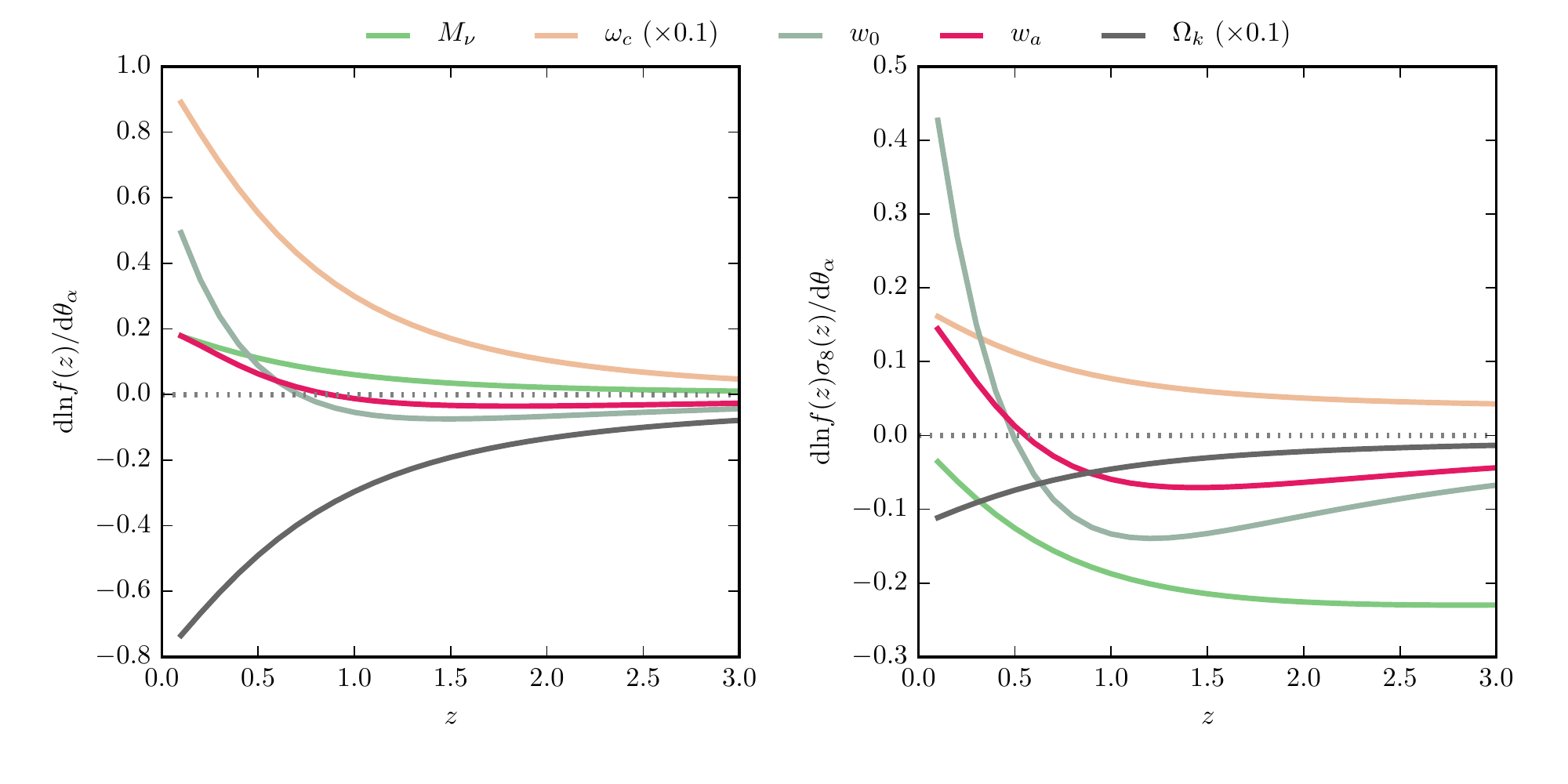}
\caption{The derivatives of $f(z)$ and $f(z)\sigma_{8}(z)$ with respect to the parameters $M_{\nu}$, $\omega_{\textrm{cdm}}$, $w_{0}$, $w_{a}$ and $\Omega_{k}$. The effects of adding extensions to the $\Lambda$CDM model on the neutrino mass constraints can be understood by comparing these derivatives. The values of $f$ used are those for $k=10^{-3} h$ Mpc$^{-1}$. The derivatives with respect to $\omega_{cdm}$ and $\Omega_{k}$ have been re-scaled by a factor of 0.1 for plotting purposes.}\label{fig_derivatives_growth}
\end{figure}

We briefly deconstruct the constraints we see in the left-hand panel of Figure \ref{fig_f}. The left panel of Figure \ref{fig_derivatives_growth} demonstrates the effect of changing the most relevant cosmological parameters on the large-scale value of $f$. Adding information on the large-scale structure growth rate to the CMB prior provides an improvement in the neutrino mass constraints by inverting the correlation between $\omega_{cdm}$ and $M_{\nu}$ in the CMB prior, as both paramters increase $f$ by adding additional matter. 

Increasing $w_{0}$ or $w_{a}$ reduces $f$ at higher redshifts and increases it at lower redshifts, so these two parameters become correlated with $M_{\nu}$ at the redshifts covered by our surveys, weakening the neutrino mass constraint. The effect of freeing $\Omega_{k}$ is quite complex. It can be seen from Figure \ref{fig_derivatives_growth} that the effects of $\omega_{cdm}$ and $\Omega_{k}$ on $f$ are strongly degenerate. We may expect $M_{\nu}$ and $\Omega_{k}$ to be correlated because the former increases $f$ and the latter reduces it. However, $\omega_{cdm}$ is strongly anti-correlated with $M_{\nu}$, as discussed previously, and is much more strongly correlated with $\Omega_{k}$ than $M_{\nu}$ is. So the net effect, including the distance prior, is an anti-correlation between $M_{\nu}$ and $\Omega_{k}$. This results in the weakening of constraints with free curvature seen in Figure \ref{fig_f}. In general, we marginalise over $P_{\textrm{m}}$, $b$, $H(z)$ and $D_{A}(z)$ here. Fixing $H(z)$ and $D_{A}(z)$ in this case can actually improve constraints quite significantly (from 0.24 eV to 0.17 eV for Euclid in the $\Lambda$CDM case). This demonstrates the complementarity of BAO and RSD information.

\begin{figure}
\includegraphics[width=\textwidth]{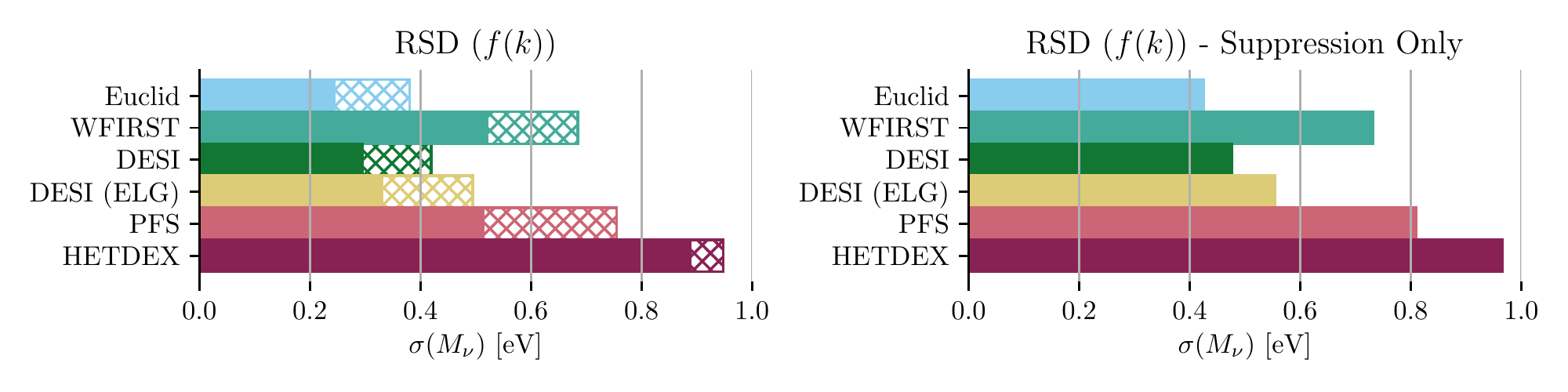}
\caption{As Figure \ref{fig_f}, but this time for all surveys. The constraints are shown here for two cosmological models - solid bars represent the $\Lambda$CDM model and the cross-hatched bars give the constraints for the most complex model (+$\Omega_{k}$, $w_{0}$, $w_{a}$).}\label{fig_f_surveys}
\end{figure}

Figure \ref{fig_f_surveys} summarises the constraints on $M_{\nu}$ from all surveys for both the simplest and most complex cosmological model. 

We can next examine the constraints achievable from the combination $f(k)\sigma_{8}$. The right panel of Figure \ref{fig_derivatives_growth} shows how $f\sigma_{8}$ varies with the key cosmological parameters. In this case, the choice of whether to include information on $A_s\exp(-2\tau)$ and $\tau$ in the CMB prior plays a crucial role in the constraints on $M_{\nu}$ obtained. 

\begin{figure}
\includegraphics[width=\textwidth]{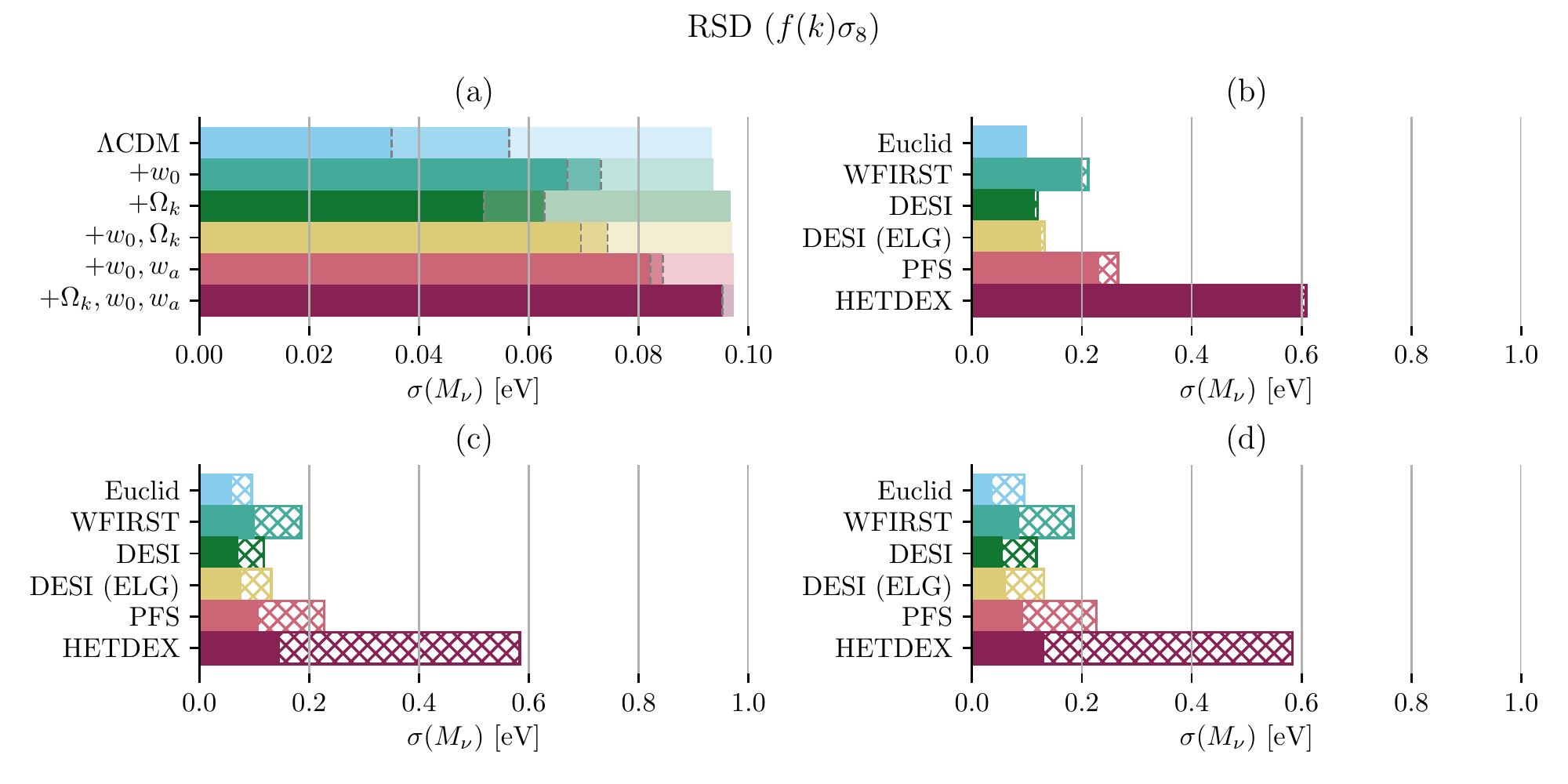}
\caption{A breakdown of the forecasted constraints on $M_{\nu}$ from RSD for various surveys, priors and cosmological models. In this case, RSD is used to constrain $f\sigma_{8}$ (the shape of the matter power spectrum is marginalised over). Panel (a) shows a breakdown of the constraints by cosmological model for Euclid, with the different opacities representing different prior conditions on $A_{s}\exp(-2\tau)$ and $\tau$ (from most to least transparent: no CMB information on $A_{s}\exp(-2\tau)$ and $\tau$, Planck priors on $A_{s}\exp(-2\tau)$ and $\tau$, and Planck priors on $A_{s}\exp(-2\tau)$ with perfectly-known $\tau$). Note that the length of the x-axis has been altered here for greater precision. The three other panels summarise the same information for the other surveys, with solid bars representing $\Lambda$CDM constraints and cross-hatched bars representing $\Lambda$CDM + $\Omega_{k}+w_0+w_a$, as before. These panels differ in the priors assumed: (b) no CMB information on $A_{s}\exp(-2\tau)$ and $\tau$, (c) Planck priors on $A_{s}\exp(-2\tau)$ and $\tau$, and (d) Planck priors on $A_{s}\exp(-2\tau)$ with perfectly-known $\tau$.}\label{fig_fs8}
\end{figure}

We begin by considering panel (a) of Figure \ref{fig_fs8}, which shows a breakdown of the RSD constraints for Euclid with three possible choices of CMB prior, distinguished by different levels of transparency. The largest (most transparent) error bars correspond to the four-parameter CMB prior we have used so far (constraining $\theta_{s}$, $n_s$ and $\omega_b$). Constraining $\sigma_{8}$ requires a meaningful prior on $A_{s}$, which itself requires reasonable constraints on $\tau$ as $A_{s}\exp(-2\tau)$ is the parameter measured from the CMB. In the case in which $A_{s}\exp(-2\tau)$ and $\tau$ are not included in the CMB prior, the constraints on $M_{\nu}$ ultimately become controlled by the uncertainty on these two parameters, and freeing other parameters makes little difference, giving very uniform constraints across the cosmological models we consider. Including the constraining power of the CMB on $A_{s}\exp(-2\tau)$ and $\tau$ can improve the constraints significantly, as seen from the more opaque bars of panel (a) of Figure \ref{fig_fs8}. When $A_{s}$ is reasonably constrained, the constraint on $M_{\nu}$ becomes dominated by how well $\tau$ is known. The middle-opacity bars show the constraints when prior information from Planck on $\tau$ and $A_{s}\exp(-2\tau)$ is included. This corresponds to constraints of $\sigma(\ln 10^{10}A_s) \approx 0.04$ and $\sigma(\tau) \approx 0.02$ from the distance prior alone. The most opaque bars include the prior information from Planck on $\tau$ and $A_{s}\exp(-2\tau)$ but assume that $\tau$ is known perfectly. Further discussion on this topic is provided in Section \ref{subsection_priors}.


\subsubsection{The Shape of the Matter Power Spectrum}

\begin{figure}
\includegraphics[width=\textwidth]{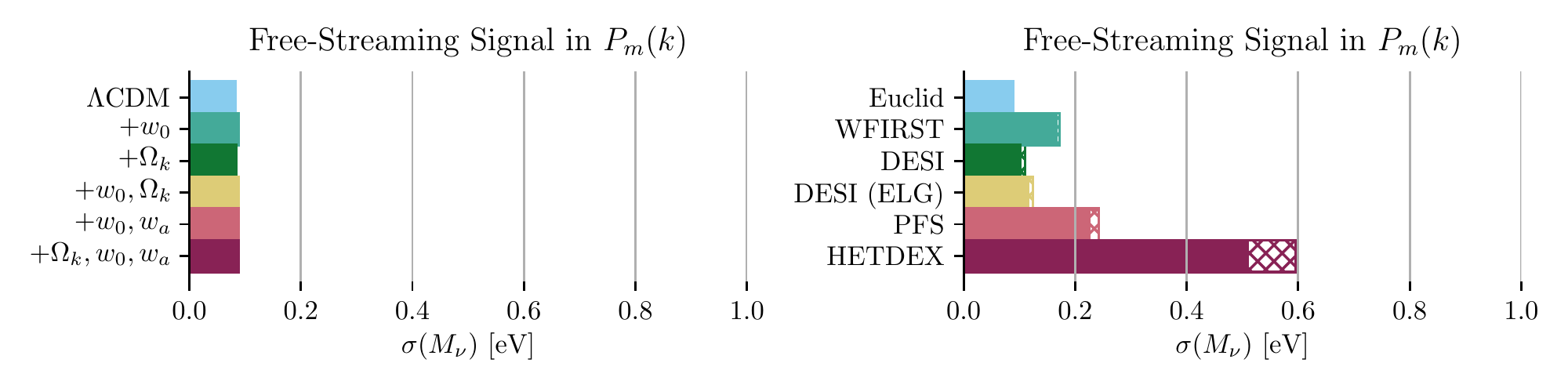}
\caption{Left panel: A breakdown of the forecasted constraints on $M_{\nu}$ for Euclid from the scale-dependent suppression of the matter power spectrum characteristic of massive neutrino free-streaming. The amplitude of the matter power spectrum at $k=10^{-3}~h$ Mpc$^{-1}$ is marginalised over. Right panel: A summary of the constraints from the same source for the surveys analysed.}\label{fig_shape}
\end{figure}

The cumulative effect of neutrinos suppressing the structure growth rate $f$ over time is a corresponding small-scale suppression in the matter power that increases in magnitude with time. We can isolate this effect by taking the power spectrum amplitude outside the derivative and smoothing out the BAO signal, to give a clear signal like that in Figure \ref{fig_derivatives}. Figure \ref{fig_shape} shows the constraints obtained from this signal alone. We see that the constraints are relatively robust against variations in the model, as we would expect. We also note that these constraints are tighter than those from the scale-dependence of $f(k)$, as the fractional change in the matter power spectrum caused by massive neutrinos at low redshifts is larger than the fractional change in the growth factor. 

In the HETDEX case, the signal is weak and the CMB prior therefore makes up a greater component of the total constraint. When $\Omega_{k}$ is allowed to vary, the constraints from the prior on $\omega_{cdm}$ (which is strongly correlated with $M_{\nu}$) become much weaker. Therefore, for HETDEX, there is some degradation of the constraint when expanding to the most complex model.

\subsection{Combining the Suppression Signals}

\begin{figure}
\includegraphics[width=\textwidth]{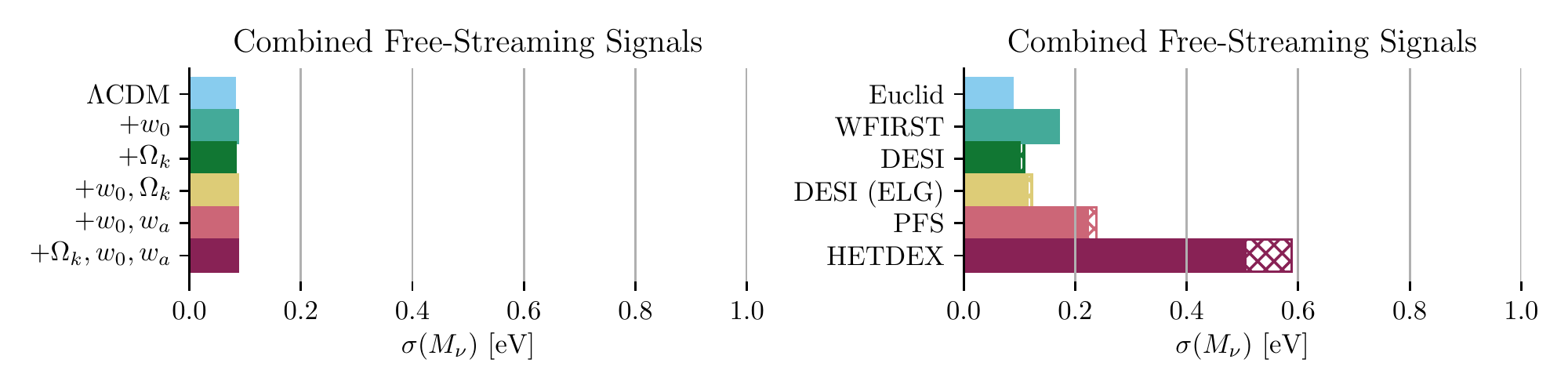}
\caption{This figure combines the constraints from the free-streaming signals in both $f(k)$ and $P_{\textrm{m}}(k)$ for Euclid. We see through comparison with Figure \ref{fig_shape} that the constraint is dominated by information from the shape of the matter power spectrum.}\label{fig_scale_dependence}
\end{figure}

The suppression of $P_{\textrm{m}}(k)$ and $f(k)$ caused by massive neutrinos on small scales can be combined to maximise the constraint from this effect. The result is dominated by the shape of $P_{\textrm{m}}$, and information from the effect on $f$ is a much more minor contribution, but does provide some improvement on the constraints from $P_{\textrm{m}}$ alone. Figure \ref{fig_scale_dependence} demonstrates this. This combination is the most robust probe of the neutrino mass that we identify in this work. As Figure \ref{fig_scale_dependence} demonstrates, the constraints are not dependent on basic assumptions about the dark energy equation of state or curvature. The constraints provided by this combination are also competitive with constraints from distance probes and RSD.

\subsection{Combined Information}

\begin{figure}
\includegraphics[width=\textwidth]{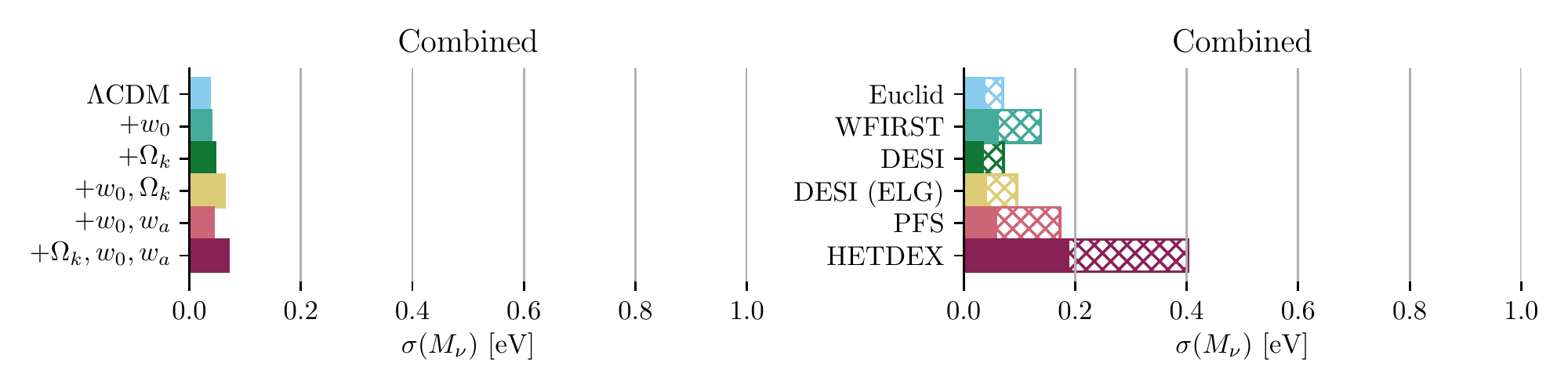}
\caption{The total combined neutrino mass constraint forecasts for Euclid (left) and for all surveys (right). The cross-hatched bars show the degradation in the constraint if curvature and the dark energy equation of state are allowed to vary. No CMB prior on $A_{s}\exp(-2\tau)$ or $\tau$ is included.}\label{fig_combined}
\end{figure}

Ultimately, the constraints from the total galaxy power spectrum can be broken down into two categories - distance constraints (BAO, AP, etc.) and constraints from the growth of structure (the shape and amplitude of the matter power spectrum, RSD, etc.). Combining these two probes to extract the maximum amount of information is powerful. Figure \ref{fig_combined} shows a breakdown in the combined constraints for Euclid, without any CMB prior being included for $A_{s}\exp(-2\tau)$ and $\tau$ (the effect of adding these priors is demonstrated in Section \ref{subsection_priors}). We see that the constraints suffer considerably if more model parameters are allowed to vary. For example, assuming $\Lambda$CDM with Euclid gives a constraint of 0.037 eV. This weakens to 0.07 eV for our most complex model. This emphasises the inherent weakness of taking constraints from the entire observed galaxy power spectrum without closer analysis. We emphasise that the constraints derived here are somewhat larger than other published results. This is because of the choice of the compressed likelihood prior (see Section \ref{subsection_method_priors}), which constrains only $\theta_s$, $n_s$ and $\omega_{b}$, and does not include the constraining power of CMB lensing (see Section \ref{subsection_choice_of_cmb_prior} for discussion).

\section{Discussion}\label{section_discussion}

\subsection{Significance}

In the previous section, we isolated the neutrino mass information available from the distinctive effects of neutrino free-streaming on $P_{\textrm{m}}(k)$ and $f(k)$ in galaxy surveys. The constraints from this signal alone are weaker than the combined constraints that are usually quoted. However, the probes we have emphasised are directly relateable to neutrino physics and are relatively insensitive to the assumed cosmological model, unlike in the combined case. In a time when the $\Lambda$CDM model still leaves many open questions, an upper limit on the neutrino mass can only be asserted with confidence if the constraint is reasonably independent of small changes in the assumptions about the underlying cosmology. The unique effects of massive neutrinos on the growth of structure provide this. Our calculations therefore provide conservative constraints that should be convincing to particle physicists and those working outside the cosmology community. 

There are effects beyond the cosmological extensions that we have considered here that also induce a scale-dependence in the structure growth rate, for example, modified gravity models such as $f(R)$ theories (a study of the degeneracies between $f(R)$ gravity and massive neutrinos was carried out in \cite{baldi_cosmic_2014}). However, we consider it unlikely that the characteristic scale and magnitude of the suppression caused by massive neutrinos would be exactly replicated by another effect, particularly if the redshift-dependence of the effect could be measured.

\subsection{Choice of CMB prior}\label{subsection_choice_of_cmb_prior}

Our forecasted constraints are somewhat weaker than those in other published work because of our choice of a very conservative CMB prior, which neglects some important effects that can help constrain $M_{\nu}$. The most significant of these is CMB lensing, which probes the shape and amplitude of the matter power spectrum directly, and is considered in \cite{wu_guide_2014, abazajian_neutrino_2015, allison_towards_2015, pan_constraints_2015, archidiacono_physical_2017}. Other smaller effects of neutrino mass that are neglected include an early ISW effect and changes to the diffusion damping scale (see \cite{archidiacono_physical_2017} for a summary). As our aim in this work has been to disentangle neutrino mass constraints, using a very simple prior made sense, and for testing the cosmological dependence of different probes, we also needed a cosmology-independent prior. A similar deconstruction of how CMB lensing complements each of the types of information isolated in this paper is currently in the works.

\subsection{Sensitivity to priors on $A_{s}$ and $\tau$}\label{subsection_priors}

The small-scale CMB temperature power spectrum is sensitive to the parameter combination $A_{s}\exp(-2\tau)$ \cite{planck_collaboration_planck_2014}, making $A_{s}$ and $\tau$ very strongly correlated when using CMB temperature anisotropy data alone. Strengthening the constraint on $\tau$ is often specified as a recommended route towards improving constraints on $M_{\nu}$ \cite[e.g.][]{allison_towards_2015, liu_eliminating_2016}. The constraints on the neutrino mass from structure growth information (parameterised, for example, by $f\sigma_{8}$) are sensitive to the primordial amplitude $A_{s}$, but not sensitive to $\tau$. However, because $A_{s}$ and $\tau$ are so strongly correlated in the CMB prior, adding the prior makes the constraint on $M_{\nu}$ strongly dependent on the weak constraint on $\tau$ provided by CMB polarisation. Ultimately, when constraints are strong enough, the $\tau$ constraint becomes the limiting factor when trying to strengthen the constraint on $M_{\nu}$. This is why we can obtain significant improvements in some of our results by extending our compressed likelihood CMB prior to also include constraints on $A_{s}\exp(-2\tau)$ and $\tau$. We can expect the weak constraints on $\tau$ from Planck to be significantly improved by 21 cm emission measurements used to probe the epoch of reionisation \cite{liu_eliminating_2016}, and we provide results for fixed $\tau$ to give forecasts in the most optimistic cases.

\begin{figure}
\includegraphics[width=0.75\textwidth]{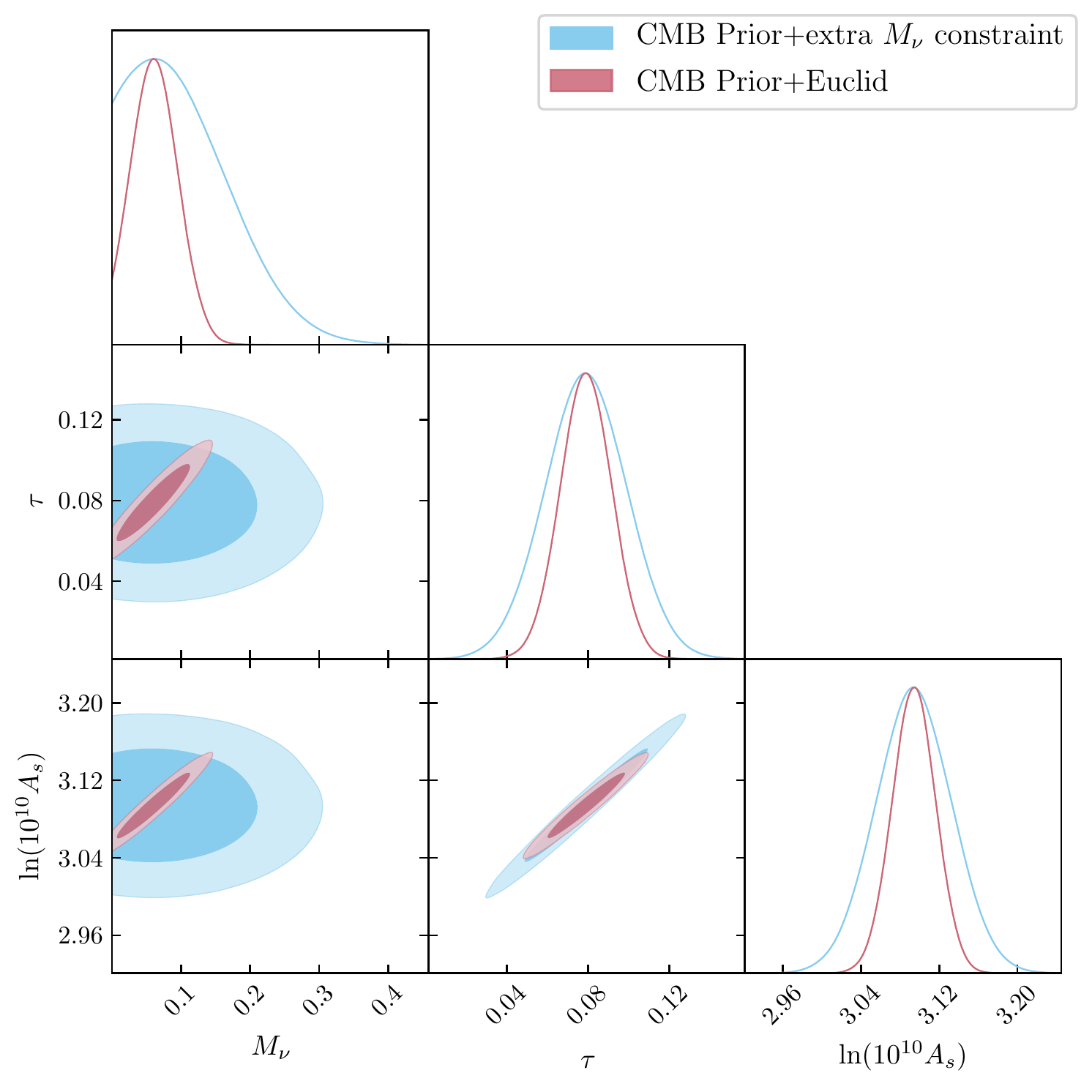}
\caption{Contour plot demonstrating how the constraints on $M_{\nu}$ from the CMB prior and the combined galaxy redshift survey information become dominated by the constraints on $\tau$ when Planck priors on $A_{s}\exp(-2\tau)$ and $\tau$ are included. Note that in the CMB prior only case, an \lq extra $M_{\nu}$ constraint\rq~is included to keep the 1-$\sigma$ error on $M_{\nu}$ below 0.1 eV. This is purely for demonstrative purposes - without this the CMB prior would only impose an error of 1.0 eV (see Table \ref{Table_priors}), which would inflate the axis scales and make the contour for the CMB prior combined with Euclid ($\sigma M_{\nu} \approx$ 0.03 eV) difficult to make out. We see that when the CMB prior alone is used, there is little correlation between $M_{\nu}$ and either $A_{s}$ or $\tau$, but the latter two are strongly correlated with each other. When the combined information from the Euclid survey is added, $M_{\nu}$ becomes strongly correlated with $A_{s}$, making it also strongly correlated with $\tau$. $\tau$ is currently only weakly constrained by CMB polarisation information, so the neutrino mass constraint becomes limited by our knowledge of the value of $\tau$. }\label{fig_contour_tau}
\end{figure}

\begin{figure}
\includegraphics[width=\textwidth]{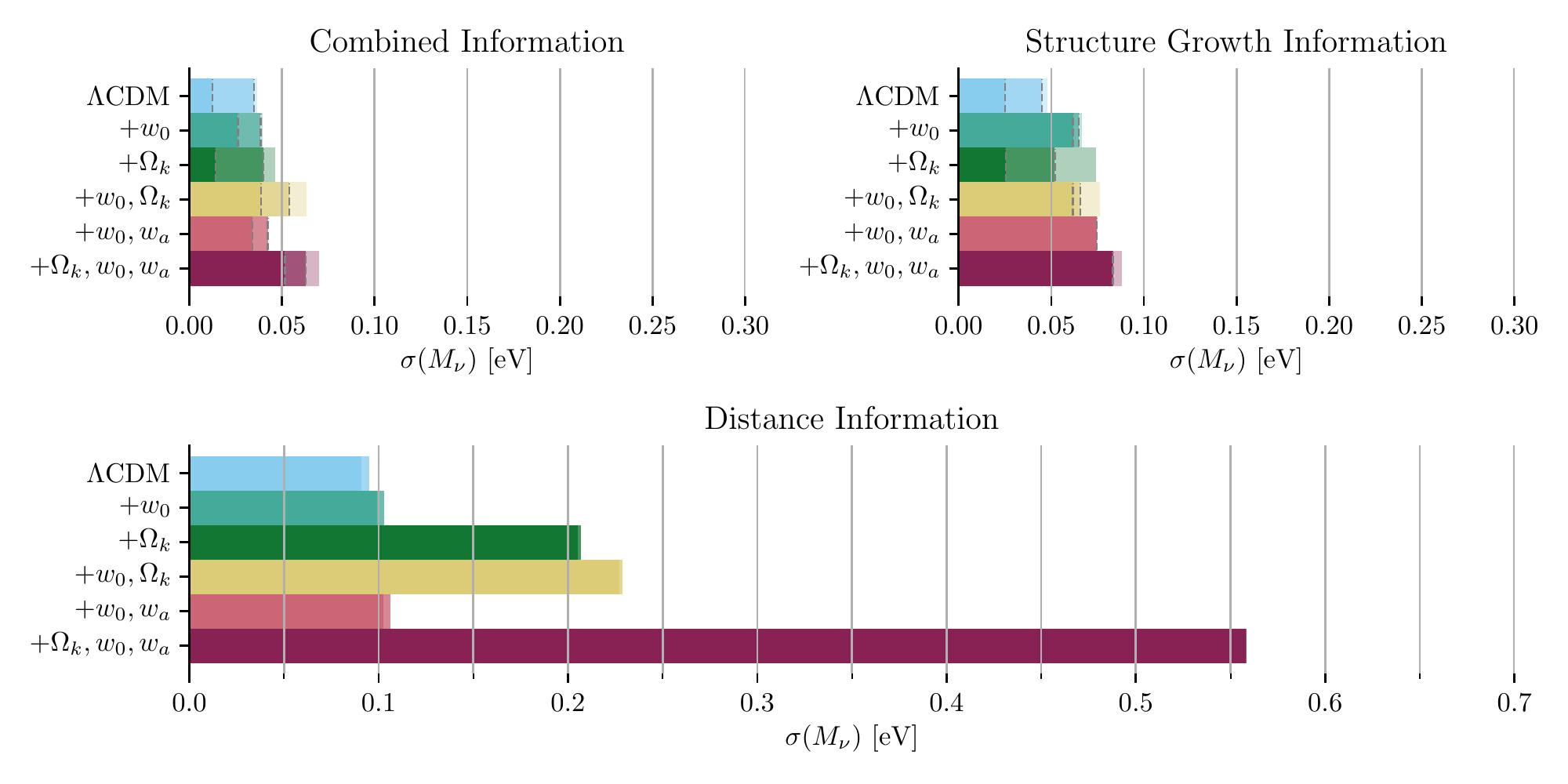}
\caption{The forecasted constraints from Euclid, with the combined constraints broken down into their broad components, for various choices of prior for $A_{s}\exp(-2\tau)$ and $\tau$. The different opacities represent these different priors (from most to least transparent: no CMB information on $A_{s}\exp(-2\tau)$ and $\tau$, Planck priors on $A_{s}\exp(-2\tau)$ and $\tau$, and Planck priors on $A_{s}\exp(-2\tau)$ with perfectly-known $\tau$). The x-axis of the distance information panel is extended to keep roughly the same scale as the other two panels for comparison.}\label{fig_summary}
\end{figure}

In Figure \ref{fig_contour_tau}, we show how the neutrino mass constraint is limited by the constraint on $\tau$ when $\tau$ is included in the CMB prior. In Figure \ref{fig_summary}, we demonstrate the impact on the neutrino mass constraints from Euclid  when $\tau$ is perfectly known. As previously noted, distance measurements have no dependence on $A_{s}$, and are therefore unaffected by our choice to improve constraints on these parameters. But in the case of structure growth probes there is a significant improvement, which ultimately leads to an improvement in the combined constraints, particularly for the simpler models. 

\subsection{Non-Linear Effects}

The minimum scale chosen for most of our calculations ($k_{\textrm{max}}=0.2~h~\textrm{Mpc}^{-1}$) is a scale at which non-linear effects may play a small role. However, the primary aim of the present work is to examine the relative strength of the different probes we describe, and we expect that all of these constraints would be modified similarly by the inclusion of non-linearities. We have also run our results for $k_{\textrm{max}}=0.15~h~\textrm{Mpc}^{-1}$ and find that there is not a significant qualitative change in our conclusions.

Additionally, in principle, there is more information on the neutrino mass available in the non-linear regime, as non-linearity makes the suppression of the galaxy power spectrum due to free-streaming more prominent \cite{saito_impact_2008}. We have decided not to include that information in our calculations, and our forecast is therefore conservative.

\subsection{Strengthening the neutrino signal}

There are a number of possible approaches that could be taken to enhance the strength of the neutrino mass constraint available from the robust signals from galaxy surveys identified and isolated in this work - the suppression of $P_{\textrm{m}}(z,k)$ and $f(z,k)$ on small scales. The most obvious aspiration would be to improve our understanding of non-linearities in the power spectrum, which would allow us to increase $k_{\textrm{max}}$ further into the regime where these effects are most obvious. 

We could also extend to other cosmological experiments. The effects of free-streaming on the power spectrum caused by massive neutrinos on small scales can also be probed through weak lensing and CMB lensing. A similar deconstruction to that applied to galaxy surveys in this paper may be in order to isolate this information source within lensing. Massive neutrinos also introduce a scale-dependence in the linear halo bias (see \cite{loverde_neutrino_2016, loverde_halo_2014, loverde_neutrino_2014}). This is another distinctive massive neutrino signal that could be incorporated into this work. All of these will be a focus of future work.

\section{Conclusions}\label{section_conclusions}

In this paper, we have analysed the various components of the observed galaxy power spectrum as will be measured by future galaxy redshift surveys to determine how these different components can contribute to a determination of the total neutrino mass, $M_{\nu}$. Adding massive neutrinos to a cosmological model alters the expansion rate (an effect that can be identified through measurements of $H(z)$ and $D_{A}(z)$ derived from the AP test and standard rulers such as the BAO signal). Massive neutrinos also modify the structure growth rate (an effect that can be measured using RSD). The most unique identifier of massive neutrinos, however, is a small-scale relative suppression in the matter power spectrum and in the structure growth rate, as the free-streaming of massive neutrinos reduces clustering on small scales. This is a probe of the neutrino mass that provides constraints that are independent of simple extensions to the assumed cosmological model. In this paper, we have disentangled all of these measurement tools, and demonstrated the sensitivity of each to the assumed cosmological model. We have used a minimalistic CMB prior (constraining $R$, $l_{A}$, $\omega_{b}$ and $n_{s}$), which provides constraints that are relatively insensitive to the assumed curvature or dark energy equation of state.

We have also provided forecasts of the neutrino mass constraint using all of the information available in the observed galaxy power spectrum combined. These constraints can be weakened by more than a factor of two when the curvature or dark energy parameters are allowed to vary. We have confirmed that the combined constraints on $M_{\nu}$ are limited by the accuracy to which $\tau$ is known \cite{allison_towards_2015}. 

To extract the most robust constraints possible on $M_{\nu}$, future surveys should focus on constraining $P_{\textrm{m}}(k)$ and $f(k)$ as precisely as possible, which requires solid measurements of the broadband shape of the galaxy power spectrum and redshift-space distortion analysis that considers the scale-dependence of $f$. Large-scale and small-scale values of $P_{\textrm{m}}$ and $f$ can be compared to attempt to identify the relative suppression of structure growth caused by massive neutrino free-streaming. As the magnitudes of these two effects varies somewhat with redshift (see Figure \ref{fig_derivatives}), surveys with deeper redshift surveys may provide more definitive results by showing a contrast between the signals measured at different redshifts.

\acknowledgments

This work was supported in part by MEXT KAKENHI Grant Number 15H05896. We are very grateful to Ryu Makiya and Chi-Ting Chiang for helpful suggestions and for providing results for comparison. We would also like to thank Francesco Villaescusa-Navarro for his useful comments.

\clearpage

\appendix
\section{Extended Results}\label{app_extended_results}

Here we provide a summary of constraints from various sources and surveys for the $\Lambda$CDM model. An impression of how these should results vary with extensions to the cosmological model can be obtained using the figures in the main text.

Note that the \lq Combined\rq~constraints provided here are on the large side compared to other numbers in the literature. This is because of our very conservative choice of CMB prior (see Section \ref{subsection_method_priors} and Section \ref{subsection_choice_of_cmb_prior} for discussion). 

\begin{table}[h!]
\begin{tabular}{lrrrrrr}
\hline
\textbf{Various: $\bf{\Lambda}$CDM}                           &   Euclid &   WFIRST &   DESI &   DESI (ELG) &   PFS &   HETDEX \\
\hline
 Distance Full             &     0.1  &     0.13 &   0.11 &         0.12 &  0.16 &     0.38 \\
 RSD ($f(k)$)              &     0.24 &     0.52 &   0.29 &         0.33 &  0.51 &     0.89 \\
 Combined Scale Dependence &     0.08 &     0.16 &   0.1  &         0.11 &  0.22 &     0.5  \\
\hline
\end{tabular}
\caption{Forecasted constraints on $M_{\nu}$ for each survey assuming $\Lambda$CDM, using the full distance information, constraints on $f(k)$ extracted from RSD, and the combination of the free-streaming signals in the matter power spectrum shape and the structure growth rate.}
\end{table}

\begin{table}[h!]

\begin{tabular}{lrrrrrr}
\hline
\textbf{Combined: $\bf{\Lambda}$CDM}                                        &   Euclid &   WFIRST &   DESI &   DESI (ELG) &   PFS &   HETDEX \\
\hline
 No CMB $A_s/\tau$ constraint          &     0.04 &     0.06 &   0.03 &         0.04 &  0.06 &     0.19 \\
 Planck $A_s/\tau$ constraint          &     0.03 &     0.05 &   0.03 &         0.04 &  0.05 &     0.11 \\
 Planck $A_s$ constraint; $\tau$ fixed &     0.01 &     0.02 &   0.01 &         0.02 &  0.03 &     0.09 \\
\hline
\end{tabular}
\caption{Forecasted constraints on $M_{\nu}$ for each survey assuming $\Lambda$CDM, using the full combined information, with various choices of priors for $A_{s}$ and $\tau$.}
\end{table}

\begin{table}[h!]

\begin{tabular}{lrrrrrr}
\hline
\textbf{RSD ($\bf{f(k)\sigma_8}$): $\bf{\Lambda}$CDM}                                       &   Euclid &   WFIRST &   DESI &   DESI (ELG) &   PFS &   HETDEX \\
\hline
 No CMB $A_s/\tau$ constraint          &     0.09 &     0.2  &   0.11 &         0.12 &  0.23 &     0.6  \\
 Planck $A_s/\tau$ constraint          &     0.06 &     0.1  &   0.07 &         0.07 &  0.11 &     0.14 \\
 Planck $A_s$ constraint; $\tau$ fixed &     0.04 &     0.08 &   0.05 &         0.06 &  0.09 &     0.13 \\
\hline
\end{tabular}
\caption{Forecasted constraints on $M_{\nu}$ for each survey assuming $\Lambda$CDM, using all of the available RSD information (constraints on $f(k)\sigma_{8}$, with various choices of priors for $A_{s}$ and $\tau$.}
\end{table}

\section{Choice of Power Spectrum}\label{app_choice_of_power_spectrum}

One subtlety that was pointed out to us after the submission of our article is that the galaxy power spectrum (Eq. \ref{eq_full_pk}) should more accurately be given as:

\begin{equation}
P_g(k, \mu) = \left[ b+f_{cb}(k)\mu^2\right]^{2}P_{cb}(k) + \bar{n}_{g}^{-1},
\end{equation}

\noindent where $P_{cb}$ is the power spectrum of baryons and cold dark matter only, and $f_{cb}$ is the growth rate corresponding to this power spectrum. The motivation for this is that neutrino perturbations do not contribute to the formation of galaxy haloes. We examined the effect of this change on some of our results for Euclid and found some change in our constraints, but not sufficient to significantly change our conclusions. The combined constraints are degraded by about 10\%. The constraints from distance measurements are unaffected and those from RSD (full $f\sigma_8$, including scale-dependence of $f$) are actually improved by about 10\%. The constraints from the combined free-streaming signature are degraded by less than 20\%.

\bibliographystyle{JHEP}

\bibliography{kingfisher}

\providecommand{\href}[2]{#2}\begingroup\raggedright\begin{thebibliography}{10}

\bibitem{capozzi_neutrino_2016}
F.~Capozzi, E.~Lisi, A.~Marrone, D.~Montanino and A.~Palazzo, \emph{Neutrino
  masses and mixings: {Status} of known and unknown \$3{\textbackslash}nu\$
  parameters},
  \href{http://dx.doi.org/10.1016/j.nuclphysb.2016.02.016}{\emph{Nuclear
  Physics B} {\bf 908} (July, 2016) 218--234}.

\bibitem{planck_collaboration_planck_2016}
P.~Collaboration, P.~A.~R. Ade, N.~Aghanim, M.~Arnaud, M.~Ashdown, J.~Aumont
  et~al., \emph{Planck 2015 results. {XIII}. {Cosmological} parameters},
  \href{http://dx.doi.org/10.1051/0004-6361/201525830}{\emph{Astronomy \&
  Astrophysics} {\bf 594} (Oct., 2016) A13}.

\bibitem{abazajian_cmb-s4_2016}
K.~N. Abazajian, P.~Adshead, Z.~Ahmed, S.~W. Allen, D.~Alonso, K.~S. Arnold
  et~al., \emph{{CMB}-{S}4 {Science} {Book}, {First} {Edition}},
  {\emph{arXiv:1610.02743 [astro-ph, physics:gr-qc, physics:hep-ph,
  physics:hep-th]} (Oct., 2016) }.

\bibitem{hu_weighing_1998}
W.~Hu, D.~J. Eisenstein and M.~Tegmark, \emph{Weighing {Neutrinos} with
  {Galaxy} {Surveys}},
  \href{http://dx.doi.org/10.1103/PhysRevLett.80.5255}{\emph{Physical Review
  Letters} {\bf 80} (June, 1998) 5255--5258}.

\bibitem{takada_cosmology_2006}
M.~Takada, E.~Komatsu and T.~Futamase, \emph{Cosmology with {High}-redshift
  {Galaxy} {Survey}: {Neutrino} {Mass} and {Inflation}},
  \href{http://dx.doi.org/10.1103/PhysRevD.73.083520}{\emph{Physical Review D}
  {\bf 73} (Apr., 2006) 083520}.

\bibitem{wu_guide_2014}
W.~L.~K. Wu, J.~Errard, C.~Dvorkin, C.~L. Kuo, A.~T. Lee, P.~McDonald et~al.,
  \emph{A {Guide} to {Designing} {Future} {Ground}-based {CMB} {Experiments}},
  \href{http://dx.doi.org/10.1088/0004-637X/788/2/138}{\emph{The Astrophysical
  Journal} {\bf 788} (June, 2014) 138}.

\bibitem{abazajian_neutrino_2015}
K.~N. Abazajian, K.~Arnold, J.~Austermann, B.~A. Benson, C.~Bischoff, J.~Bock
  et~al., \emph{Neutrino {Physics} from the {Cosmic} {Microwave} {Background}
  and {Large} {Scale} {Structure}},
  \href{http://dx.doi.org/10.1016/j.astropartphys.2014.05.014}{\emph{Astroparticle
  Physics} {\bf 63} (Mar., 2015) 66--80}.

\bibitem{allison_towards_2015}
R.~Allison, P.~Caucal, E.~Calabrese, J.~Dunkley and T.~Louis, \emph{Towards a
  cosmological neutrino mass detection},
  \href{http://dx.doi.org/10.1103/PhysRevD.92.123535}{\emph{Physical Review D}
  {\bf 92} (Dec., 2015) 123535}.

\bibitem{pan_constraints_2015}
Z.~Pan and L.~Knox, \emph{Constraints on neutrino mass from {Cosmic}
  {Microwave} {Background} and {Large} {Scale} {Structure}},
  \href{http://dx.doi.org/10.1093/mnras/stv2164}{\emph{Monthly Notices of the
  Royal Astronomical Society} {\bf 454} (Dec., 2015) 3200--3206}.

\bibitem{archidiacono_physical_2017}
M.~Archidiacono, T.~Brinckmann, J.~Lesgourgues and V.~Poulin, \emph{Physical
  effects involved in the measurements of neutrino masses with future
  cosmological data},
  \href{http://dx.doi.org/10.1088/1475-7516/2017/02/052}{\emph{JCAP} {\bf 1702}
  (Feb., 2017) 052}.

\bibitem{eisenstein_detection_2005}
D.~J. Eisenstein, I.~Zehavi, D.~W. Hogg, R.~Scoccimarro, M.~R. Blanton, R.~C.
  Nichol et~al., \emph{Detection of the {Baryon} {Acoustic} {Peak} in the
  {Large}-{Scale} {Correlation} {Function} of {SDSS} {Luminous} {Red}
  {Galaxies}}, \href{http://dx.doi.org/10.1086/466512}{\emph{Astrophys.J.} {\bf
  633} (2005) 560--574}.

\bibitem{cole_2df_2005}
S.~Cole, W.~J. Percival, J.~A. Peacock, P.~Norberg, C.~M. Baugh, C.~S. Frenk
  et~al., \emph{The 2df {Galaxy} {Redshift} {Survey}: {Power}-spectrum analysis
  of the final dataset and cosmological implications},
  \href{http://dx.doi.org/10.1111/j.1365-2966.2005.09318.x}{\emph{Mon.Not.Roy.Astron.Soc.}
  {\bf 362} (2005) 505--534}.

\bibitem{alcock_evolution_1979}
C.~Alcock and B.~Paczynski, \emph{An evolution free test for non-zero
  cosmological constant},
  \href{http://dx.doi.org/10.1038/281358a0}{\emph{Nature} {\bf 281} (1979)
  358--359}.

\bibitem{kaiser_clustering_1987}
N.~Kaiser, \emph{Clustering in real space and in redshift space},
  {\emph{Mon.Not.Roy.Astron.Soc.} {\bf 227} (1987) 1--27}.

\bibitem{lesgourgues_massive_2006}
J.~Lesgourgues and S.~Pastor, \emph{Massive neutrinos and cosmology},
  \href{http://dx.doi.org/10.1016/j.physrep.2006.04.001}{\emph{Physics Reports}
  {\bf 429} (July, 2006) 307--379}.

\bibitem{hernandez_neutrino_2017}
O.~F. Hern\'{a}ndez, \emph{Neutrino {Masses}, {Scale}-{Dependent} {Growth}, and
  {Redshift}-{Space} {Distortions}},
  \href{http://dx.doi.org/10.1088/1475-7516/2017/06/018}{\emph{JCAP} {\bf 1706}
  (June, 2017) 018}.

\bibitem{planck_collaboration_planck_2016-1}
P.~Collaboration, P.~A.~R. Ade, N.~Aghanim, M.~Arnaud, M.~Ashdown, J.~Aumont
  et~al., \emph{Planck 2015 results. {XIV}. {Dark} energy and modified
  gravity},
  \href{http://dx.doi.org/10.1051/0004-6361/201525814}{\emph{Astronomy \&
  Astrophysics} {\bf 594} (Oct., 2016) A14}.

\bibitem{qian_neutrino_2015}
X.~Qian and P.~Vogel, \emph{Neutrino {Mass} {Hierarchy}},
  \href{http://dx.doi.org/10.1016/j.ppnp.2015.05.002}{\emph{Progress in
  Particle and Nuclear Physics} {\bf 83} (July, 2015) 1--30}.

\bibitem{percival_large_2013}
W.~J. Percival, \emph{Large {Scale} {Structure} {Observations}},
  {\emph{arXiv:1312.5490 [astro-ph]} (Dec., 2013) }.

\bibitem{font-ribera_desi_2014}
A.~Font-Ribera, P.~McDonald, N.~Mostek, B.~A. Reid, H.-J. Seo and A.~Slosar,
  \emph{{DESI} and other dark energy experiments in the era of neutrino mass
  measurements},
  \href{http://dx.doi.org/10.1088/1475-7516/2014/05/023}{\emph{Journal of
  Cosmology and Astroparticle Physics} {\bf 2014} (May, 2014) 023--023}.

\bibitem{blas_cosmic_2011}
D.~Blas, J.~Lesgourgues and T.~Tram, \emph{The {Cosmic} {Linear} {Anisotropy}
  {Solving} {System} ({CLASS}) {II}: {Approximation} schemes},
  \href{http://dx.doi.org/10.1088/1475-7516/2011/07/034}{\emph{Journal of
  Cosmology and Astroparticle Physics} {\bf 2011} (July, 2011) 034--034}.

\bibitem{eisenstein_robustness_2007}
D.~J. Eisenstein, H.-j. Seo and M.~White, \emph{On the {Robustness} of the
  {Acoustic} {Scale} in the {Low}-{Redshift} {Clustering} of {Matter}},
  \href{http://dx.doi.org/10.1086/518755}{\emph{The Astrophysical Journal} {\bf
  664} (Aug., 2007) 660--674}.

\bibitem{seo_improved_2007}
H.-J. Seo and D.~J. Eisenstein, \emph{Improved forecasts for the baryon
  acoustic oscillations and cosmological distance scale},
  \href{http://dx.doi.org/10.1086/519549}{\emph{The Astrophysical Journal} {\bf
  665} (Aug., 2007) 14--24}.

\bibitem{eisenstein_improving_2007}
D.~J. Eisenstein, H.-j. Seo, E.~Sirko and D.~Spergel, \emph{Improving
  {Cosmological} {Distance} {Measurements} by {Reconstruction} of the {Baryon}
  {Acoustic} {Peak}}, \href{http://dx.doi.org/10.1086/518712}{\emph{The
  Astrophysical Journal} {\bf 664} (Aug., 2007) 675--679}.

\bibitem{eisenstein_baryonic_1998}
D.~J. Eisenstein and W.~Hu, \emph{Baryonic {Features} in the {Matter}
  {Transfer} {Function}}, \href{http://dx.doi.org/10.1086/305424}{\emph{The
  Astrophysical Journal} {\bf 496} (Apr., 1998) 605--614}.

\bibitem{savitzky_smoothing_1964}
A.~Savitzky and M.~J.~E. Golay, \emph{Smoothing and {Differentiation} of {Data}
  by {Simplified} {Least} {Squares} {Procedures}.},
  \href{http://dx.doi.org/10.1021/ac60214a047}{\emph{Analytical Chemistry} {\bf
  36} (July, 1964) 1627--1639}.

\bibitem{shoji_extracting_2009}
M.~Shoji, D.~Jeong and E.~Komatsu, \emph{Extracting {Angular} {Diameter}
  {Distance} and {Expansion} {Rate} of the {Universe} from {Two}-dimensional
  {Galaxy} {Power} {Spectrum} at {High} {Redshifts}: {Baryon} {Acoustic}
  {Oscillation} {Fitting} versus {Full} {Modeling}},
  \href{http://dx.doi.org/10.1088/0004-637X/693/2/1404}{\emph{The Astrophysical
  Journal} {\bf 693} (Mar., 2009) 1404--1416}.

\bibitem{white_forecasting_2009}
M.~White, Y.-S. Song and W.~J. Percival, \emph{Forecasting {Cosmological}
  {Constraints} from {Redshift} {Surveys}},
  \href{http://dx.doi.org/10.1111/j.1365-2966.2008.14379.x}{\emph{Monthly
  Notices of the Royal Astronomical Society} {\bf 397} (Aug., 2009)
  1348--1354}.

\bibitem{hill_hobby-eberly_2008}
G.~J. Hill, K.~Gebhardt, E.~Komatsu, N.~Drory, P.~J. MacQueen, J.~Adams et~al.,
  \emph{The {Hobby}-{Eberly} {Telescope} {Dark} {Energy} {Experiment}
  ({HETDEX}): {Description} and {Early} {Pilot} {Survey} {Results}},
  {\emph{arXiv:0806.0183 [astro-ph]} (June, 2008) }.

\bibitem{leung_bayesian_2017}
A.~S. Leung, V.~Acquaviva, E.~Gawiser, R.~Ciardullo, E.~Komatsu, G.~R. Zeimann
  et~al., \emph{Bayesian {Redshift} {Classification} of {Emission}-line
  {Galaxies} with {Photometric} {Equivalent} {Widths}},
  \href{http://dx.doi.org/10.3847/1538-4357/aa71af}{\emph{Astrophys.J.} {\bf
  843} (July, 2017) 130}.

\bibitem{takada_extragalactic_2014}
M.~Takada, R.~Ellis, M.~Chiba, J.~E. Greene, H.~Aihara, N.~Arimoto et~al.,
  \emph{Extragalactic {Science}, {Cosmology} and {Galactic} {Archaeology} with
  the {Subaru} {Prime} {Focus} {Spectrograph} ({PFS})},
  \href{http://dx.doi.org/10.1093/pasj/pst019}{\emph{Publications of the
  Astronomical Society of Japan} {\bf 66} (Feb., 2014) R1}.

\bibitem{desi_collaboration_desi_2016}
D.~Collaboration, A.~Aghamousa, J.~Aguilar, S.~Ahlen, S.~Alam, L.~E. Allen
  et~al., \emph{The {DESI} {Experiment} {Part} {I}: {Science},{Targeting}, and
  {Survey} {Design}}, {\emph{arXiv:1611.00036 [astro-ph]} (Oct., 2016) }.

\bibitem{amendola_cosmology_2016}
L.~Amendola, S.~Appleby, A.~Avgoustidis, D.~Bacon, T.~Baker, M.~Baldi et~al.,
  \emph{Cosmology and {Fundamental} {Physics} with the {Euclid} {Satellite}},
  {\emph{arXiv:1606.00180 [astro-ph]} (June, 2016) }.

\bibitem{laureijs_euclid_2011}
R.~Laureijs, J.~Amiaux, S.~Arduini, J.-L. Auguères, J.~Brinchmann, R.~Cole
  et~al., \emph{Euclid {Definition} {Study} {Report}}, {\emph{arXiv:1110.3193
  [astro-ph]} (Oct., 2011) }.

\bibitem{spergel_wide-field_2013}
D.~Spergel, N.~Gehrels, J.~Breckinridge, M.~Donahue, A.~Dressler, B.~S. Gaudi
  et~al., \emph{Wide-{Field} {InfraRed} {Survey} {Telescope}-{Astrophysics}
  {Focused} {Telescope} {Assets} {WFIRST}-{AFTA} {Final} {Report}},
  {\emph{arXiv:1305.5422 [astro-ph]} (May, 2013) }.

\bibitem{green_wide-field_2012}
J.~Green, P.~Schechter, C.~Baltay, R.~Bean, D.~Bennett, R.~Brown et~al.,
  \emph{Wide-{Field} {InfraRed} {Survey} {Telescope} ({WFIRST}) {Final}
  {Report}}, {\emph{arXiv:1208.4012 [astro-ph]} (Aug., 2012) }.

\bibitem{baldi_cosmic_2014}
M.~Baldi, F.~Villaescusa-Navarro, M.~Viel, E.~Puchwein, V.~Springel and
  L.~Moscardini, \emph{Cosmic {Degeneracies} {I}: {Joint} {N}-body
  {Simulations} of {Modified} {Gravity} and {Massive} {Neutrinos}},
  \href{http://dx.doi.org/10.1093/mnras/stu259}{\emph{Monthly Notices of the
  Royal Astronomical Society} {\bf 440} (May, 2014) 75--88}.

\bibitem{planck_collaboration_planck_2014}
P.~Collaboration, P.~A.~R. Ade, N.~Aghanim, C.~Armitage-Caplan, M.~Arnaud,
  M.~Ashdown et~al., \emph{Planck 2013 results. {XVI}. {Cosmological}
  parameters},
  \href{http://dx.doi.org/10.1051/0004-6361/201321591}{\emph{Astronomy \&
  Astrophysics} {\bf 571} (Nov., 2014) A16}.

\bibitem{liu_eliminating_2016}
A.~Liu, J.~R. Pritchard, R.~Allison, A.~R. Parsons, U.~Seljak and B.~D.
  Sherwin, \emph{Eliminating the optical depth nuisance from the {CMB} with 21
  cm cosmology},
  \href{http://dx.doi.org/10.1103/PhysRevD.93.043013}{\emph{Physical Review D}
  {\bf 93} (Feb., 2016) }.

\bibitem{saito_impact_2008}
S.~Saito, M.~Takada and A.~Taruya, \emph{{Impact of massive neutrinos on
  nonlinear matter power spectrum}},
  \href{http://dx.doi.org/10.1103/PhysRevLett.100.191301}{\emph{Phys. Rev.
  Lett.} {\bf 100} (2008) 191301}, [\href{https://arxiv.org/abs/0801.0607}{{\tt
  0801.0607}}].

\bibitem{loverde_neutrino_2016}
M.~LoVerde, \emph{Neutrino mass without cosmic variance},
  \href{http://dx.doi.org/10.1103/PhysRevD.93.103526}{\emph{Physical Review D}
  {\bf 93} (May, 2016) 103526}.

\bibitem{loverde_halo_2014}
M.~LoVerde, \emph{Halo bias in mixed dark matter cosmologies},
  \href{http://dx.doi.org/10.1103/PhysRevD.90.083530}{\emph{Physical Review D}
  {\bf 90} (Oct., 2014) 083530}.

\bibitem{loverde_neutrino_2014}
M.~LoVerde and M.~Zaldarriaga, \emph{Neutrino clustering around spherical dark
  matter halos},
  \href{http://dx.doi.org/10.1103/PhysRevD.89.063502}{\emph{Physical Review D}
  {\bf 89} (Mar., 2014) 063502}.

\end{thebibliography}\endgroup

\end{document}